\documentclass[10pt,journal,twoside]{IEEEtran}

\usepackage[cmex10]{amsmath}
\usepackage[caption=false,font=footnotesize]{subfig}
\usepackage{graphicx, amssymb, xcolor, cite, booktabs, multirow, url, pgfplots, bm, bbm}

\newtheorem{proposition}{Proposition}
\newtheorem{lemma}{Lemma}
\newtheorem{corollary}{Corollary}

\newcommand{\eps}{\varepsilon}

\newcommand{\expn}{\mathbb{E}}
\newcommand{\pr}{\mathbb{P}}
\newcommand{\pmark}{\mathcal{P}_{\text{mark}}}
\renewcommand{\vec}{\bm}

\newcommand{\thint}{\vec{\theta}_{\text{int}}}
\newcommand{\thall}{\vec{\theta}_{\text{all1}}}
\newcommand{\thmaj}{\vec{\theta}_{\text{maj}}}
\newcommand{\thmin}{\vec{\theta}_{\text{min}}}
\newcommand{\thcoi}{\vec{\theta}_{\text{coin}}}
\newcommand{\thadd}{\vec{\theta}_{\text{add}}}
\newcommand{\thdil}{\vec{\theta}_{\text{dil}}}
\newcommand{\ththr}{\vec{\theta}_{\text{thr}}}

\newcommand{\tos}{^s}
\newcommand{\caps}{C\tos}
\newcommand{\pints}{p_{\text{int}}\tos}
\newcommand{\palls}{p_{\text{all1}}\tos}
\newcommand{\pmajs}{p_{\text{maj}}\tos}
\newcommand{\pmins}{p_{\text{min}}\tos}
\newcommand{\pcois}{p_{\text{coin}}\tos}
\newcommand{\padds}{p_{\text{add}}\tos}
\newcommand{\pdils}{p_{\text{dil}}\tos}

\newcommand{\toj}{^j}
\newcommand{\capj}{C\toj}

\newcommand{\pallj}{p_{\text{all1}}\toj}
\newcommand{\pmajj}{p_{\text{maj}}\toj}
\newcommand{\pminj}{p_{\text{min}}\toj}
\newcommand{\pcoij}{p_{\text{coin}}\toj}
\newcommand{\paddj}{p_{\text{add}}\toj}
\newcommand{\pdilj}{p_{\text{dil}}\toj}
\newcommand{\pthrj}{p_{\text{thr}}\toj}

\begin{document}


\title{Asymptotics of Fingerprinting and Group Testing: Tight Bounds from Channel Capacities}
\author{Thijs~Laarhoven%
\thanks{T.~Laarhoven is with the Department of Mathematics and Computer Science, Eindhoven University of Technology, The Netherlands.\protect\\
E-mail: {\ttfamily mail@thijs.com}.\protect\\%
Part of the material in this paper will be presented at the 2nd ACM Workshop on Information Hiding and Multimedia Security (Salzburg, Austria, June 2014).}}

\maketitle



\begin{abstract}
In this work we consider the large-coalition asymptotics of various fingerprinting and group testing games, and derive explicit expressions for the capacities for each of these models. We do this both for simple decoders (fast but suboptimal) and for joint decoders (slow but optimal). 

For fingerprinting, we show that if the pirate strategy is known, the capacity often decreases linearly with the number of colluders, instead of quadratically as in the uninformed fingerprinting game. For many attacks the joint capacity is further shown to be strictly higher than the simple capacity. 

For group testing, we improve upon known results about the joint capacities, and derive new explicit asymptotics for the simple capacities. These show that existing simple group testing algorithms are suboptimal, and that simple decoders cannot asymptotically be as efficient as joint decoders. For the traditional group testing model, we show that the gap between the simple and joint capacities is a factor $\log_2(e) \approx 1.44$ for large numbers of defectives.
\end{abstract}


\IEEEpeerreviewmaketitle


\section{Introduction}
\label{sec:intro}


\subsection{Fingerprinting}
\label{sec:intro-fp}

\IEEEPARstart{T}{o} protect copyrighted content against unauthorized redistribution, distributors commonly embed watermarks or fingerprints in the content, uniquely linking copies to individual users. If the distributor finds an illegal copy of the content online, he can then extract the watermark from this copy and compare it to the database of watermarks, to determine which user was responsible. 

To combat this solution, a group of $c$ pirates may try to form a coalition and perform a collusion attack. By comparing their unique versions of the content, they will detect differences in their copies which must be part of the watermark. They can then try to create a mixed pirate copy, where the resulting watermark matches the watermark of different pirates in different segments of the content, making it hard for the distributor to find the responsible users. The goal of the distributor of the content is to assign the watermarks to the users in such a way that, even if many pirates collude, the pirate copy can still be traced back to the responsible users.


\subsection{Group testing}
\label{sec:intro-gt}

A different area of research that has received considerable attention in the last few decades is group testing, introduced by Dorfman~\cite{dorfman43} in the 1940s. Suppose a large population contains a small number $c$ of infected (or defective) items. To identify these items, it is possible to perform group tests: testing a subset of the population will lead to a positive test result if this subset contains at least one defective item, and a negative result otherwise. Since the time to run a single test may be very long, the subsets to test need to be chosen in advance, after which all group tests are performed simultaneously. Then, when the test results come back, the subset of defective items needs to be identified. The goal of the game is to identify these defectives using as few group tests as possible, and with a probability of error as small as possible. 


\subsection{Model}
\label{sec:intro-model}

The above problems of fingerprinting and group testing can be jointly modeled by the following two-person game between (in terms of fingerprinting) the distributor $\mathcal{D}$ and the adversary $\mathcal{C}$ (the set of colluders, or the set of defectives). Throughout the paper we will mostly use terminology from fingerprinting (i.e. users instead of items, colluders instead of defective items), unless we are specifically dealing with group testing results.

First, there is a universe $\mathcal{U}$ of $n$ users, and the adversary is assigned a random subset of users $\mathcal{C} \subseteq \mathcal{U}$ of size $|\mathcal{C}| = c$. This subset $\mathcal{C}$ is unknown to the distributor (but we assume that the distributor does know the size $c$ of $\mathcal{C}$), and the aim of the game for the distributor is ultimately to discover $\mathcal{C}$. The two-person game consists of three phases: (1) the distributor uses an \textit{encoder} to generate a fingerprinting code, used for assigning versions to users; (2) the colluders employ a \textit{collusion channel} to generate the pirate output from their given code words; and (3) the distributor uses a \textit{decoder} to map the pirate output to a set $\mathcal{C}' \subseteq \mathcal{U}$.


\subsubsection{Encoder}

First, the distributor generates a fingerprinting code $\mathcal{X}$ of $n$ binary code words of length $\ell$.\footnote{In fingerprinting a common generalization is to assume that the entries of the code words come from an alphabet of size $q \geq 2$, but in this paper we restrict our attention to the binary case $q = 2$.} The parameter $\ell$ is referred to as the code length, and the distributor would like $\ell$ to be as small as possible. For the eventual embedded watermark, we assume that for each segment of the content there are two differently watermarked versions, so the watermark of user $j$ is determined by the $\ell$ entries in the $j$th code word of $\mathcal{X}$. 

A common restriction on the encoding process is to assume that $\mathcal{X}$ is created by first generating a bias vector $\vec{P} \in (0,1)^{\ell}$ (by choosing each entry $P_i$, for $i = 1, \dots, \ell$, independently from a certain distribution $f_P$), and then generating code words $\vec{X}_j \in \mathcal{X}$ according to $\pr(X_{j,i} = 1) = P_i$. This guarantees that watermarks of different users $j$ are independent, and that watermarks in different positions $i$ are independent. Fingerprinting schemes that satisfy this assumption are sometimes called bias-based schemes, and the encoders in this paper (both for group testing and fingerprinting) are also assumed to belong to this category.


\subsubsection{Collusion channel}

After generating $\mathcal{X}$, the code words are used to select and embed watermarks in the content, and the content is sent out to all users. The colluders then get together, compare their copies, and use a certain collusion channel or pirate attack $\vec{\Theta}$ to determine the pirate output $\vec{Y} \in \{0,1\}^{\ell}$. If the pirate attack behaves symmetrically both in the colluders and in the positions $i$, then the collusion channel can be modeled by a vector $\vec{\theta} \in [0,1]^{c+1}$, consisting of entries $\theta_z = \pr(Y_i = 1|z)$ (for $z = 0, \dots, c$) indicating the probability of outputting a $1$ when the pirates received $z$ ones and $c - z$ zeroes. A further restriction on $\vec{\theta}$ in fingerprinting is the marking assumption introduced by Boneh and Shaw~\cite{boneh98}, which says that $\theta_0 = 0$ and $\theta_c = 1$, i.e., if the pirates receive only zeros or ones they have to output this symbol.


\subsubsection{Decoder}

Finally, after the pirate output has been generated and distributed, we assume the distributor intercepts it and applies a decoding algorithm to $\vec{Y}$, $\mathcal{X}$ and $\vec{P}$ to compute a set $\mathcal{C}' \subseteq \mathcal{U}$ of accused users. The distributor wins the game if $\mathcal{C}' = \mathcal{C}$ and loses if this is not the case.\footnote{In this paper we thus consider the catch-all scenario, where not at least one colluder (the catch-one scenario) but \textit{all} colluders should be found for the distributor to win the game.}


\subsubsection*{Fingerprinting vs. group testing}

While the above model is described in fingerprinting terminology, it also covers many common group testing models. The users then correspond to items, the colluders translate to defectives, the code $\mathcal{X}$ corresponds to the group testing matrix $X$ (where $X_{j,i} = 1$ if item $j$ is included in the $i$th test), and the pirate output corresponds to positive/negative test results. The collusion channel is exactly what separates group testing from fingerprinting: while in fingerprinting it is commonly assumed that this channel is not known or only weakly known to the distributor, in group testing this channel is usually assumed known in advance. This means that there is no malicious adversary in group testing, but only a randomization procedure that determines $\vec{Y}$. Note also that in (noisy) group testing, the Boneh-Shaw marking assumption may not always hold.


\subsection{Related work}
\label{sec:intro-related}

Work on the fingerprinting game described above started in the late 90s, and lower bounds on the code length were established of the order $\ell \propto c \ln n$~\cite{boneh98}, until in 2003 Tardos~\cite{tardos03} proved a lower bound of the order $\ell \propto c^2 \ln n$ and described a scheme with $\ell = O(c^2 \ln n)$, showing this bound is tight. The lower bound however did not explicitly mention leading constants, so later work on fingerprinting focused on finding this constant. Amiri and Tardos~\cite{amiri09} and Huang and Moulin~\cite{huang09, huang09b, huang10, huang12, moulin08} independently worked on this problem and found that the exact asymptotic lower bound on $\ell$ is $\ell \sim 2 c^2 \ln n$ for large $n$ and $c$. Huang and Moulin further derived the pirate strategy and encoder achieving this lower bound, which were later used~\cite{laarhoven14ihmmsec, meerwald11b, oosterwijk13b} to construct efficient schemes matching these bounds. Most of this work on lower bounds considers the setting with worst-case pirate attacks, and little is known about lower bounds for specific, suboptimal pirate attacks considered in e.g.~\cite{berchtold12, charpentier09, furon09b, laarhoven13wifs, meerwald11b, oosterwijk13b}. It is well known that for suboptimal pirate attacks the required code length may be significantly smaller than the $\ell \sim 2 c^2 \ln n$ for arbitrary attacks, but no tight bounds are known.

Research on the group testing problem started much longer ago, and already in 1985 exact asymptotics on the code length for probabilistic schemes were derived as $\ell \sim c \log_2 n$~\cite{sebo85}, whereas deterministic schemes require a code length of $\ell \propto c^2 \ln n$~\cite{dyachkov82, dyachkov89}. Later work focused on slight variations of the classical model such as noisy group testing, where a positive result may not always correspond to the presence of a defective item due to `noise' in the test output~\cite{atia09, atia12, cheraghchi09, cheraghchi11, hwang76b, laarhoven13allerton, sejdinovic10}, and threshold group testing, where the test result may only be positive if sufficiently many defective items are included in the tested subset~\cite{ahlswede11, ahlswede11b, ahlswede13, chen09, cheraghchi13, damaschke06, laarhoven13allerton, lebedev10}. For noisy and threshold group testing, exact asymptotics on the capacities are yet unknown, and so it is not known whether existing constructions are optimal.


\subsection{Contributions and outline}
\label{sec:intro-contributions}

In this paper we extend the work of Huang and Moulin~\cite{huang12} to various fingerprinting and group testing models where $\vec{\theta}$ is known in advance. We derive the simple capacities (Section~\ref{sec:cap-simple}) and the joint capacities (Section~\ref{sec:cap-joint}) for these pirate attacks/group testing models, and prove that these results are the exact optima. An overview of these results can be found in Table~\ref{tab:tab1}. Finally, we discuss the results (Section~\ref{sec:discussion}) and mention some directions for future work (Section~\ref{sec:openproblems}). 

\begin{table*}[!t]
\renewcommand{\arraystretch}{1.3}
\caption{An overview of the capacity results derived in this paper. The top rows describe pirate attacks in fingerprinting and the bottom rows consider various different models in group testing. The results for the interleaving attack and unknown attacks~\cite{huang12} and the joint capacity for the classical group testing model~\cite{sebo85} were derived before, while the other results are new. \label{tab:tab1}}
\centering
\begin{tabular}{clr@{$/$}l@{$\approx$}r@{$/$}lcr@{$/$}l@{$\approx$}r@{$/$}lc} \toprule
& Attacks / Models & \multicolumn{4}{c}{\quad Simple capacities} & (Section~\ref{sec:cap-simple}) & \multicolumn{4}{c}{Joint capacities} & (Section~\ref{sec:cap-joint}) \\ \midrule
\parbox[t]{2mm}{\multirow{6}{*}{\rotatebox[origin=c]{90}{Fingerprinting}}} & $\thint$: interleaving attack & $\left(\frac{1}{2 \ln 2}\right)$ & $c^2$ \ & \ $0.72$ & $c^2$ & (Section~\ref{sec:cap-simple-int}) %
  & $\left(\frac{1}{2 \ln 2}\right)$ & $c^2$ \ & \ $0.72$ & $c^2$ & (Section~\ref{sec:cap-joint-int}) \\
& $\thall$: all-$1$ attack & $(\ln 2)$ & $c$ & $0.69$ & $c$ & (Section~\ref{sec:cap-simple-all}) %
  & $(1)$ & $c$ & $1.00$ & $c$ & (Section~\ref{sec:cap-joint-all}) \\
& $\thmaj$: majority voting & $\left(\frac{1}{\pi \ln 2}\right)$ & $c$ & $0.46$ & $c$ & (Section~\ref{sec:cap-simple-maj}) %
  & $(1)$ & $c$ & $1.00$ & $c$ & (Section~\ref{sec:cap-joint-maj}) \\
& $\thmin$: minority voting & $(\ln 2)$ & $c$ & $0.69$ & $c$ & (Section~\ref{sec:cap-simple-min}) %
  & $(1)$ & $c$ & $1.00$ & $c$ & (Section~\ref{sec:cap-joint-min}) \\
& $\thcoi$: coin-flip attack & $\left(\frac{1}{4}\ln 2\right)$ & $c$ & $0.17$ & $c$ & (Section~\ref{sec:cap-simple-coi})%
  & $\left(\log_2(\tfrac{5}{4})\right)$ & $c$ & $0.32$ & $c$ & (Section~\ref{sec:cap-joint-coi}) \\
& $\pmark$: unknown attacks & $\left(\frac{1}{2 \ln 2}\right)$ & $c^2$ & $0.72$ & $c^2$ & (Section~\ref{sec:cap-simple-unk}) %
  & $\left(\frac{1}{2 \ln 2}\right)$ & $c^2$ & $0.72$ & $c^2$ & (Section~\ref{sec:cap-joint-unk}) \\ \midrule
\parbox[t]{2mm}{\multirow{6}{*}{\rotatebox[origin=c]{90}{Group testing}}} & $\thall$: classical model & $(\ln 2)$ & $c$ & $0.69$ & $c$ & (Section~\ref{sec:cap-simple-ord}) %
  & $(1)$ & $c$ & $1.00$ & $c$ & (Section~\ref{sec:cap-joint-ord}) \\
& $\thadd$: additive noise & $(\ln 2 - r)$ & $c$ & $0.69$ & $c$ & (Section~\ref{sec:cap-simple-add}) %
  & $(1 - \frac{1}{2} h(r))$ & $c$ & $1.00$ & $c$ & (Section~\ref{sec:cap-joint-add}) \\
& $\thdil$: dilution noise & $(\ln 2 - O(r \ln r))$ & $c$ & $0.69$ & $c$ & (Section~\ref{sec:cap-simple-dil}) %
  & \ \ $(1 - \frac{1}{2} h(r) \ln 2)$ & $c$ & $1.00$ & $c$ & (Section~\ref{sec:cap-joint-dil}) \\ 
& $\ththr^{(u)}$: threshold (no gap) & \multicolumn{4}{c}{between $0.46/c$ and $0.69/c$} & (Section~\ref{sec:cap-simple-th1}) %
  & $(1)$ & $c$ & $1.00$ & $c$ & (Section~\ref{sec:cap-joint-th1}) \\
& $\thint^{(l,u)}$: threshold (int.\ gap) & \multicolumn{4}{c}{between $0.72/c^2$ and $0.69/c$} & (Section~\ref{sec:cap-simple-th2}) %
  & \multicolumn{4}{c}{\ \ \ between $0.72/c^2$ and $1.00/c$} & (Section~\ref{sec:cap-joint-th2}) \\
& $\thcoi^{(l,u)}$: threshold (coin.\ gap) & \multicolumn{4}{c}{between $0.17/c$ and $0.69/c$} & (Section~\ref{sec:cap-simple-th2}) %
  & \multicolumn{4}{c}{\ \ \ between $0.32/c$ and $1.00/c$} & (Section~\ref{sec:cap-joint-th2}) \\ \bottomrule
\end{tabular}
\end{table*}


\section{Simple capacities}
\label{sec:cap-simple}

In simple decoding, \textit{``the receiver makes an innocent/guilty decision on each user independently of the other users, and there lies the simplicity but also the suboptimality of this decoder."}~\cite[Section 4.3]{moulin08} In other words, a simple decoder bases its decision whether or not to accuse user $j$ only on the $j$th code word of $\mathcal{X}$, and not on other code words in $\mathcal{X}$. This means that the decoding step will generally be fast but less accurate than when all information available to the decoder (the entire code $\mathcal{X}$) is taken into account. 

Huang and Moulin~\cite{huang09, huang09b, huang10, huang12} previously studied simple capacities in the context of fingerprinting, and showed that given a set of allowed collusion channels $\mathcal{P}^c$ (depending only on $\vec{\theta}$) and a set of allowed encoders $\mathcal{P}^e$, any fingerprinting rate below
\begin{align}
\caps(\mathcal{P}^e, \mathcal{P}^c) = \max_{f_P \in \mathcal{P}^e} \min_{\vec{\theta} \in \mathcal{P}^c} \expn_P I(p, \vec{\theta}) \label{eq:caps}
\end{align}
is achievable\footnote{Certain conditions on $\mathcal{P}^c$ and $\mathcal{P}^e$ need to be satisfied for this to hold, but we refer the reader to \cite{huang12} for details.}, where
\begin{align}
I(p, \vec{\theta}) = I(X_1; Y | P = p)
\end{align}
is the mutual information between a colluder's symbol $X_1$ and the pirate output $Y$ in one segment $i$. In this work we will always let $\mathcal{P}^e$ be the set of all probability distribution functions on $(0,1)$, and we will commonly omit the argument $\mathcal{P}^e$ from $\caps$. For fixed collusion channels $\mathcal{P}^c = \{\vec{\theta}\}$, fixing $P$ is optimal~\cite[Section~IV.B]{huang12}, in which case the expression from~\eqref{eq:caps} reduces to
\begin{align}
\caps(\{\vec{\theta}\}) = \max_{f_P} \expn_P I(p, \theta) = \max_{p \in (0,1)} I(X_1; Y|P = p).
\end{align}
With slight abuse of notation we will abbreviate the left hand side as $\caps(\vec{\theta})$. We will also consider some cases where (part of) the collusion channel is unknown, in which case the capacity includes a minimization over $\vec{\theta}$ as well:
\begin{align}
\caps(\mathcal{P}^c) = \max_{f_P} \min_{\vec{\theta} \in \mathcal{P}^c} \expn_P I(X_1; Y|P = p).
\end{align}
To study the mutual information payoff function $I(p, \vec{\theta})$ we will use the following identity~\cite[Equation~(61)]{huang12}:
\begin{align}
I(p, \vec{\theta}) &= p d(a_1 \| a) + (1 - p) d(a_0 \| a),
\end{align}
where $a, a_0, a_1$ are defined as
\begin{align}
a &= \sum_{z = 0}^c \binom{c}{z} p^z (1 - p)^{c-z} \theta_z, \\
a_0 &= \sum_{z = 0}^{c-1} \binom{c - 1}{z} p^z (1 - p)^{c-z-1} \theta_z, \\
a_1 &= \sum_{z = 1}^c \binom{c - 1}{z - 1} p^{z-1} (1 - p)^{c-z} \theta_z.
\end{align}
Here, $d(\cdot \| \cdot)$ denotes the relative entropy or Kullback-Leibler divergence, defined by $d(\alpha \| \beta) = \alpha \log_2(\tfrac{\alpha}{\beta}) + (1 - \alpha) \log_2(\tfrac{1 - \alpha}{1 - \beta})$. Given $p$ and $\vec{\theta}$, the above formulas allow us to compute the mutual information $I(p, \vec{\theta})$ explicitly. 

For obtaining the simple capacities for various models, we will extensively work with the Kullback-Leibler divergence. In general analyzing this function is not so pretty, but we can simplify some computations with the following Taylor expansion around $\alpha = \beta$:
\begin{align}
d(\alpha \| \beta) = \frac{(\alpha - \beta)^2}{2 \beta(1 - \beta) \ln 2} \left(1 + O\left(\frac{|\alpha - \beta|}{\beta(1 - \beta)}\right)\right). \label{eq:d}
\end{align}
Intuitively, this says that the divergence is bigger if $\alpha$ and $\beta$ are further apart, but for $\alpha$ and $\beta$ both close to $0$ or $1$ the divergence may blow up as well due to the $\beta$ and $1 - \beta$ in the denominator. In that case we have to be careful and see whether $|\alpha - \beta|$ approaches $0$ faster than $\beta$ or $1 - \beta$. A special case of~\eqref{eq:d} for $\beta = \frac{1}{2}$ and $\alpha \approx \frac{1}{2}$ is
\begin{align}
d\left(\frac{1}{2} \pm \gamma \Big\| \frac{1}{2}\right) = \frac{2\gamma^2}{\ln 2} + O(\gamma^4). \label{eq:d2}
\end{align}
Finally, if $\alpha = \frac{1}{2}$ and $\beta \approx \frac{1}{2}$, we can rewrite $d(\alpha \| \beta)$ as
\begin{align}
d\left(\frac{1}{2} \Big\| \frac{1}{2}(1 \pm \gamma)\right) = \frac{1}{2} d\left(1 \| 1 - \gamma^2\right), \label{eq:d3}
\end{align}
and regardless of $\alpha$ and $\beta$, we always have $d(\alpha \| \beta) = d(1 - \alpha \| 1 - \beta)$.


\subsection{Fingerprinting}
\label{sec:cap-simple-fp}

We will study the simple capacities for five commonly considered fingerprinting attacks, and for completeness also mention the result of Huang and Moulin regarding the simple capacity for unknown attacks. 


\subsubsection{Interleaving attack}
\label{sec:cap-simple-int}

The interleaving attack in fingerprinting (considered in e.g.~\cite{berchtold12, charpentier09, furon09b, huang12, laarhoven13wifs, oosterwijk13b}) is characterized by the coalition choosing one of its members at random, and outputting his symbol. Given $z$ members with a $1$ and $c - z$ members with a $0$, the probability of outputting a $1$ is then equal to $\frac{z}{c}$, regardless of $z$ and $c$:
\begin{align}
(\thint)_z = \frac{z}{c} \, . \qquad (0 \leq z \leq c)
\end{align}
This attack is known to be one of the strongest pirate attacks, and the capacity is proportional to $\frac{1}{c^2}$. The exact asymptotics of the simple capacity for the interleaving attack were previously derived by Huang and Moulin.

\begin{proposition} \cite[Theorem 6]{huang12}
The simple capacity for the interleaving attack is:
\begin{align}
\caps(\thint) &= \frac{1}{2c^2 \ln 2} + O\left(\frac{1}{c^4}\right) \approx \frac{0.72}{c^2},
\end{align}
and the maximizing value of $p$ is $\pints = \frac{1}{2}$.
\end{proposition}


\subsubsection{All-$1$ attack}
\label{sec:cap-simple-all}

Another commonly considered attack is the all-$1$ attack, where pirates output a $1$ whenever they can~\cite{charpentier09, laarhoven13wifs, meerwald11b, oosterwijk13b}. Due to the marking assumption they are forced to output a $0$ when they did not receive any ones, but otherwise a coalition using the all-$1$ attack will always output a $1$:
\begin{align}
(\thall)_z = \begin{cases} 0 & \text{if } z = 0; \\ 1 & \text{if } z > 0. \end{cases}
\end{align}
We will show below that this attack is significantly weaker than the interleaving attack.

\begin{proposition} \label{thm:cap-simple-all}
The simple capacity and the corresponding maximizing value of $p$ for the all-$1$ attack are:
\begin{align}
\caps(\thall) &= \frac{\ln 2}{c} + O\left(\frac{1}{c^2}\right) \approx \frac{0.69}{c}, \label{eq:all1c} \\
\palls &= \frac{\ln 2}{c} + O\left(\frac{1}{c^2}\right) \approx \frac{0.69}{c}. \label{eq:all1p}
\end{align}
\end{proposition}

\begin{IEEEproof}
First, consider $a$, $a_0$ and $a_1$. Using $\theta_z = 0$ if $z = 0$ and $\theta_z = 1$ otherwise, we get
\begin{align}
a &= \sum_{z = 0}^c \binom{c}{z} p^z (1 - p)^{c-z} \theta_z = 1 - (1 - p)^c.
\end{align}
Working out $a_0$ and $a_1$ in a similar way, we get $a_0 = 1 - (1 - p)^{c-1}$ and $a_1 = 1$. For ease of notation, let us write $s = (1 - p)^c$ and $I(p) = I(p, \vec{\thall})$, so that we get
\begin{align}
I(p) 
 &= p d\left(1 \| 1 - s\right) + (1 - p) d\left(\frac{s}{1 - p} \Big\| s\right).
\end{align}
Now, consider the second term. For large $c$, we argue that this term is small, i.e. of the order $O(\frac{1}{c^2})$, regardless of $p$:
\begin{align*}
& (1 - p) d\left(\frac{s}{1 - p} \ \Big\| \ s\right) \\
 &= -s \log_2(1 - p) + (1 - p - s) \log_2\left(1 - \frac{ps}{(1 - p)(1 - s)}\right) \\
 &\stackrel{(a)}{=} -s \log_2(1 - p) + \frac{1 - p - s}{\ln 2} \left[\frac{-ps}{(1 - p)(1 - s)} + O\left(\frac{1}{c^2}\right)\right] \\
 &\stackrel{(b)}{=} +\frac{ps}{\ln 2} - \frac{ps}{\ln 2} \left[1 - \frac{ps}{(1 - p)(1 - s)} + O\left(\frac{1}{c^2}\right)\right] \\
 &\stackrel{(c)}{=} +\frac{ps}{\ln 2} - \frac{ps}{\ln 2} + O\left(\frac{1}{c^2}\right) = O\left(\frac{1}{c^2}\right).
\end{align*}
Here $(a)$ follows from $\frac{ps}{(1-p)(1-s)} = O(\frac{1}{c})$ and $ps = O(\frac{1}{c})$ for all $p$, $(b)$ follows from $p^2 s = O(\frac{1}{c^2})$, and $(c)$ follows from $\frac{p^2 s^2}{(1-p)(1-s)} = O(\frac{1}{c^2})$ and $p^2 s = O(\frac{1}{c^2})$ for arbitrary $p$. So we are now left with:
\begin{align}
I(p) = -p \log_2(1 - s) + O\left(\frac{1}{c^2}\right).
\end{align}
For $p$ to be a global maximum we need either that $I'(p) = 0$ or $p$ should be one of the end-points $0$ or $1$. For $p \to 0,1$ we get $I(p) \to 0$, so we need to find a value $p \in (0,1)$ with $I'(p) = 0$. Writing out the remaining term and differentiating, this condition is equivalent to
\begin{align}
\frac{cps}{(1 - p)(1 - s)} = -\ln(1 - s). \label{eq:all1der}
\end{align}
Since the left hand side is $O(1)$ regardless of $p$, the right hand side must be too, so $s = 1 - o(1)$ is excluded. To exclude the case $s = o(1)$ we rewrite~\eqref{eq:all1der} to get
\begin{align}
\frac{cp}{1 - p} = \frac{1 - s}{s}\ln\left(\frac{1}{1 - s}\right).
\end{align}
Now if $s = o(1)$ then the right hand side becomes $1 - o(1)$, which implies in the left hand side that $p = \frac{1}{c} - o(\frac{1}{c})$, which implies that $s \neq o(1)$, contradicting our assumption that $s = o(1)$. So for large $c$ a maximum can only occur at $o(1) < s < 1 - o(1)$. Suppose that $s(c) \to s^* \in (0,1)$ for $c \to \infty$, with $s^* \neq s^*(c)$ not depending on $c$. Then $p(c) \to p^* = \frac{-1}{c} \ln s^*$, so the condition on $p$ and $s$ is then asymptotically equivalent to:
\begin{align}
s^* \ln s^* = (1 - s^*) \ln (1 - s^*) + O\left(\frac{1}{c}\right).
\end{align}
This has a unique solution at $s^* = \frac{1}{2} + O(\frac{1}{c})$, leading to the given values of $\palls$ and $\caps(\thall)$.
\end{IEEEproof}

In terms of code lengths, this means that any simple decoding algorithm for the all-$1$ attack requires an asymptotic number of fingerprint positions of at least $\ell \sim \frac{1}{\ln 2} c \log_2 n \approx 2.08 c \ln n$ for large $n$. This seems to contradict earlier results of~\cite{laarhoven13allerton, laarhoven13wifs}, which suggested that under a certain Gaussian assumption, only $\ell \sim 2 c \ln n$ tests are required. This apparent contradiction is caused by the fact that the Gaussian assumption in that paper is not correct in the regime of small $p$, for which those results were derived. Rigorous analysis of the scores in~\cite{laarhoven13wifs, laarhoven13allerton} shows that with that scheme, an asymptotic code length of about $\ell \approx 3 c \ln n$ is sufficient when $p \sim \frac{1}{c}\ln(2)$, which is well above the lower bound obtained above.


\subsubsection{Majority voting}
\label{sec:cap-simple-maj}

The majority voting attack~\cite{berchtold12, charpentier09, furon09b, laarhoven13wifs, meerwald12, oosterwijk13b} is characterized by the pirates choosing the symbol they have seen the most often. To avoid ambiguity, we will assume $c$ is odd, in which case the attack is given by
\begin{align}
(\thmaj)_z = \begin{cases} 0 & \text{if } z < \frac{c}{2}; \\ 1 & \text{if } z > \frac{c}{2}. \end{cases}
\end{align}
For this attack we obtain the following result.

\begin{proposition} \label{thm:cap-simple-maj}
For the majority voting attack, the simple capacity is
\begin{align}
\caps(\thmaj) &= \frac{1}{\pi c \ln 2} + O\left(\frac{1}{c^2}\right) \approx \frac{0.46}{c},
\end{align}
and the maximizing value of $p$ is $\pmajs = \frac{1}{2}$.
\end{proposition}

\begin{IEEEproof}
As mentioned before, to avoid ambiguity we focus on the case where $c = 2c' + 1$ is odd, and due to symmetry w.l.o.g.\ we may assume that $p \leq \frac{1}{2}$. First, we have:
\begin{align}
a &= \sum_{z = c' + 1}^{2c' + 1} \binom{2c' + 1}{z} p^z (1 - p)^{2c'+1-z}, \label{eq:a}
\end{align}
and $a_0$ and $a_1$ satisfy $a_0 = a + p u$ and $a_1 = a - (1 - p) u$, where $u = \binom{2c'}{c'} p^{c'} (1 - p)^{c'}$. Now if $p = O(\frac{1}{c})$, then $a_1$ and $a_0$ quickly approach $0$ leading to $I(p) = o(\frac{1}{c})$. For the remaining case $p = \omega(\frac{1}{c})$, expanding $a$ using Sanov's theorem~\cite[Theorem 11.4.1]{cover06} we get
\begin{align}
a &\sim \exp\left[(2c' + 1)\ln(2) d\left(\tfrac{1}{2} \| p\right)\right] \\
 &\sim p^{c' + \frac{1}{2}} (1 - p)^{c' + \frac{1}{2}} 2^{2c' + 1}.
\end{align}
Using Stirling's formula for the central binomial coefficient in $u$, we obtain
\begin{align}
u = \binom{2c'}{c'} p^{c'} (1 - p)^{c'} \sim \frac{2^{2c'} p^{c'} (1 - p)^{c'}}{\sqrt{\pi c'}}\, .
\end{align}
As a consequence, $\frac{u}{a} = o(1)$, and using~\eqref{eq:d} we get
\begin{align}
d(a_0 \| a) &\sim \frac{p^2 u^2}{2 \ln 2 a(1 - a)}, \\
d(a_1 \| a) &\sim \frac{(1 - p)^2 u^2}{2 \ln 2 a(1 - a)}.
\end{align}
Combining these expressions, we get
\begin{align}
I(p) &= p d(a_1 \| a) + (1 - p) d(a_0 \| a) \\
 &\sim \frac{2^{4c'} p^{2c' + 1}(1 - p)^{2c' + 1}}{2 \pi c' a (1 - a) \ln 2}. 
\end{align}
To see that this has a maximum at $p = \frac{1}{2}$, writing out the inverse of the above expression (ignoring constants) we see that, in terms of $p$,
\begin{align}
\frac{1}{I(p)} &\propto \sum_{z_1, z_2 = 0}^{c'} \binom{2c' + 1}{z_1} \binom{2c' + 1}{z_2} \left(\frac{p}{1 - p}\right)^{z_1 - z_2} \\
&= C_1 + \sum_{z_1 < z_2} C_2 \left[\left(\frac{p}{1 - p}\right)^{z_2 - z_1} + \left(\frac{1 - p}{p}\right)^{z_2 - z_1}\right] \\
&= C_1 + \sum_{z_1 < z_2} C_2 \left[2 \cosh((z_2 - z_1) \ln x)\right],
\end{align}
where $x = \frac{1-p}{p} > 1$ for $p < \frac{1}{2}$ and $x = 1$ if $p = \frac{1}{2}$, and $C_1, C_2$ are expressions that do not depend on $p$. The function between square brackets is positive and increasing in $x$ for $x \geq 1$, so it has a global minimum at $x = 1$, corresponding to $p = \frac{1}{2}$. So the maximum for $I(p)$ is attained at $p = \frac{1}{2}$, in which case $u$ satisfies
\begin{align}
u &= \frac{1}{\sqrt{\pi c/2}}\left(1 + O\left(\frac{1}{c}\right)\right).
\end{align}
To get exact asymptotics for $I(\frac{1}{2})$, we return to the expression for $I(p)$ of~\eqref{eq:d}. Since from~\eqref{eq:a} it follows that $a = \frac{1}{2}$, and both terms are identical, we obtain:
\begin{align}
I\left(\frac{1}{2}\right) &= d\left(\frac{1}{2} + \frac{1}{\sqrt{2 \pi c}}\left[1 + O\left(\frac{1}{c}\right)\right] \Big\| \ \frac{1}{2}\right).
\end{align}
Using~\eqref{eq:d2} the result then follows.
\end{IEEEproof}

This result matches the bounds obtained in~\cite{laarhoven13allerton, laarhoven13wifs}, which showed that with an almost trivial decoding algorithm one can asymptotically achieve a code length of $\ell \sim \pi c \ln n$ for large $n$ and $c$. The construction of~\cite{laarhoven13allerton, laarhoven13wifs} is thus capacity-achieving.


\subsubsection{Minority voting}
\label{sec:cap-simple-min}

As the name suggests, when pirates use the minority voting attack~\cite{berchtold12, charpentier09, furon09b, laarhoven13wifs, oosterwijk13b}, they output the symbol they have received the least often. Due to the marking assumption they are not able to output symbols they have not received, so in the binary setting the attack is defined as follows. Again, we will assume that $c$ is odd.
\begin{align}
(\thmin)_z = \begin{cases} 0 & \text{if } z = 0 \text{ or } \frac{c}{2} < z < c; \\ 1 & \text{if } z = c \text{ or } 0 < z < \frac{c}{2}. \end{cases}
\end{align}
As shown below, this attack has the same simple capacity as the all-$1$ attack.

\begin{proposition} \label{thm:cap-simple-min}
The simple capacity and the corresponding optimal value of $p$ for the minority voting attack are:
\begin{align}
\caps(\thmin) &= \frac{\ln 2}{c} + O\left(\frac{1}{c^2}\right) \approx \frac{0.69}{c}, \\
\pmins &= \frac{\ln 2}{c} + O\left(\frac{1}{c^2}\right) \approx \frac{0.69}{c}.
\end{align}
\end{proposition}

\begin{IEEEproof}
In this case the function $I(p)$ is symmetric around $p = \frac{1}{2}$, so w.l.o.g.\ we may assume $p \leq \frac{1}{2}$. For small values of $p$, minority voting is equivalent to the all-$1$ attack up to negligible order terms, while for $p \approx \frac{1}{2}$ the attack is very similar to majority voting by $\thmin \approx 1 - \thmaj$. This means that for small $p$ the mutual information payoff will be equivalent to that of the all-$1$ attack, while for $p \approx \frac{1}{2}$ we get the same values as for majority voting. Since the simple capacity for the all-$1$ attack is higher than for majority voting, the distributor should choose $p$ close to $\palls$, leading to the result. 
\end{IEEEproof}


\subsubsection{Coin-flip attack}
\label{sec:cap-simple-coi}

Instead of choosing a pirate at random and outputting his symbol (the interleaving attack), the pirates may also decide to choose a symbol at random from their set of received symbols, without paying attention to how often they received each symbol~\cite{berchtold12, furon09b, laarhoven13wifs, oosterwijk13b}. In other words, when a coalition receives both symbols, they let a fair coin-flip decide which symbol to output. This means that the collusion channel satisfies:
\begin{align}
(\thcoi)_z = \begin{cases} 0 & \text{if } z = 0; \\ \frac{1}{2} & \text{if } 0 < z < c; \\ 1 & \text{if } z = c. \end{cases}
\end{align}
This pirate attack is weaker than the interleaving attack, but stronger than the other pirate attacks considered above.

\begin{proposition} \label{thm:cap-simple-coi}
For the coin-flip attack, the simple capacity and the corresponding maximizing value of $p$ are:
\begin{align}
\caps(\thcoi) &= \frac{\ln 2}{4 c} + O\left(\frac{1}{c^2}\right) \approx \frac{0.17}{c}, \\
\pcois &= \frac{\ln 2}{2c} + O\left(\frac{1}{c^2}\right) \approx \frac{0.35}{c}.
\end{align}
\end{proposition}

\begin{IEEEproof}
Since $I(p)$ is symmetric around $p = \frac{1}{2}$, let us assume w.l.o.g.\ that $p \leq \frac{1}{2}$. For $a$, $a_0$ and $a_1$ we obtain:
\begin{align}
a &= \frac{1}{2}\left(1 + p^c - (1 - p)^c\right), \\
a_0 &= \frac{1}{2}\left(1 - (1 - p)^{c-1}\right), \\
a_1 &= \frac{1}{2}\left(1 + p^{c-1}\right).
\end{align}
So for the mutual information, we obtain
\begin{align}
& I(p) = p d\left(\frac{1}{2}\left(1 + p^{c-1}\right) \Big\| \frac{1}{2}\left(1 + p^c - (1 - p)^c\right)\right) \\
 &+ (1 - p) d\left(\frac{1}{2}\left(1 - (1 - p)^{c-1}\right) \Big\| \frac{1}{2}\left(1 + p^c - (1 - p)^c\right)\right).
\end{align}
For $p \leq \frac{1}{2}$, the terms $p^c$ and $p^{c-1}$ are negligible, so up to small order terms, we get
\begin{align}
& I(p) = p d\left(\frac{1}{2} \Big\| \frac{1}{2}\left(1 - (1 - p)^c\right)\right) \\
 &+ (1 - p) d\left(\frac{1}{2}\left(1 - (1 - p)^{c-1}\right) \Big\| \frac{1}{2}\left(1 - (1 - p)^c\right)\right).
\end{align}
Similar to the proof of the all-$1$ attack, the second term is $O(\frac{1}{c^2})$, while using~\eqref{eq:d3} we can rewrite the first term to a recognizable form:
\begin{align}
I(p) &= \frac{1}{2} \left[-p \log \left(1 - (1 - p)^{2c}\right)\right] + O\left(\frac{1}{c^2}\right).
\end{align}
The term between square brackets is exactly the dominating term for the simple capacity of the all-$1$ attack for $c' = 2c$. In other words:
\begin{align}
I_c(p, \thcoi) = \frac{1}{2} I_{2c}(p, \thall) + O\left(\frac{1}{c^2}\right).
\end{align}
Using Proposition~\ref{thm:cap-simple-all}, the result follows.
\end{IEEEproof}

For this attack, the result in~\cite{laarhoven13wifs} was also too optimistic due to the incorrect Gaussian assumption. Any simple decoder must have a code length of at least $\ell \sim \frac{4}{\ln 2} c \log_2 n \approx 8.33 c \ln n$, while the result in~\cite{laarhoven13wifs} suggests that a code length of $\ell \sim 4 c \ln n$ suffices under a certain Gaussian assumption. Again, the Gaussian assumption is to blame, and since the optimal value of $p$ is even smaller here than for the all-$1$ attack, the error of~\cite{laarhoven13wifs} is even bigger here.


\subsubsection{Unknown attacks}
\label{sec:cap-simple-unk}

Finally, the most often studied setting in fingerprinting is the scenario where the pirate attack is not known to the distributor. Due to the marking assumption the distributor does know that $\theta_0 = 0$ and $\theta_c = 1$, but otherwise no assumptions are made on the pirate strategy. The set of allowed attacks can then be described as
\begin{align}
\pmark = \{\vec{\theta} \in [0,1]^{c + 1} \mid \theta_0 = 0, \theta_c = 1\}.
\end{align}
Huang and Moulin solved the related max-min game for large $c$, and found the asymptotic optimal encoder and collusion channel leading to the saddle point solution.

\begin{proposition} \cite[Theorem 6, Corollary 7]{huang12}
The simple capacity for the uninformed fingerprinting game is
\begin{align}
\caps(\pmark) &= \frac{1}{2 c^2 \ln 2} + O\left(\frac{1}{c^3}\right) \approx \frac{0.72}{c^2},
\end{align}
and the optimizing encoder $f_P$ and collusion channel $\vec{\theta}$ achieving this bound for large $c$ are the arcsine distribution, defined by 
\begin{align} 
f_P^*(p) = \frac{1}{\pi \sqrt{p(1 - p)}}, \qquad (p \in (0,1))
\end{align}
and the interleaving attack $\thint$.
\end{proposition}


\subsection{Group testing}
\label{sec:cap-simple-gt}

For group testing, we will study five different models: the classical (noiseless) model, the models with additive noise and dilution noise, and threshold group testing with and without gaps. Other models where the test result $Y$ depends only on the tally $Z$ may be analyzed in a similar fashion.


\subsubsection{Classical model}
\label{sec:cap-simple-ord}

In the classical model, the outcome of a group test is positive iff at least one defective was present in the tested pool. This model is equivalent to the all-$1$ attack in fingerprinting, as was previously noted in e.g.~\cite{laarhoven13allerton, meerwald11b, stinson00}. This immediately leads to the following result.

\begin{corollary} \label{thm:cap-simple-cla}
For the classical group testing model, the simple informed capacity and the corresponding optimal value of $p$ are:
\begin{align}
\caps(\thall) &= \frac{\ln 2}{c} + O\left(\frac{1}{c^2}\right) \approx \frac{0.69}{c} \, , \\
\palls &= \frac{\ln 2}{c} + O\left(\frac{1}{c^2}\right) \approx \frac{0.69}{c} \, .
\end{align}
\end{corollary}

In terms of group testing algorithms, this means that any simple decoding algorithm for $c$ defectives and $n$ total items requires an asymptotic number of group tests $\ell$ of at least 
\begin{align}
\ell \sim \frac{c \log_2 n}{\ln 2} \approx 1.44 \, c \log_2 n \approx 2.08 \, c \ln n,
\end{align}
where the asymptotics are for $n \to \infty$. This improves upon the known lower bound for joint decoders of $\ell \geq c \log_2 n$ for large $n$~\cite{sebo85}, and this shows that the algorithm of Chan et al.~\cite{chan11} (which achieves a code length of $\ell \sim e \ln n$) is suboptimal. The related paper~\cite{laarhoven14decoders} shows how this bound can actually be achieved with efficient simple decoders.


\subsubsection{Additive noise}
\label{sec:cap-simple-add}

The classical group testing model is sometimes considered to be too optimistic, as the outcome of the group tests may not always be accurate. One `noisy' variant of the classical model that is sometimes considered in the literature is the additive noise model~\cite{atia12, chan11, cheraghchi11, sejdinovic10}, where a test result may even be positive (with some small probability $r$) if there were no defectives in the tested group. This corresponds to the following channel $\thadd$:
\begin{align}
(\thadd)_z = \begin{cases} r & \text{if } z = 0; \\ 1 & \text{if } z > 0. \end{cases} 
\end{align}
For small $r$ we do not expect the simple capacity or the optimal choice of $p$ to change drastically compared to the classical model, and the following analysis confirms this.

\begin{proposition} \label{thm:cap-simple-add}
For the additive noise model with parameter $r$, the simple capacity and the maximizing value of $p$ are:
\begin{align}
\caps(\thadd) &= \frac{\ln 2}{c}\left(1 - \frac{r}{\ln 2} + O(r^2)\right) + O\left(\frac{1}{c^2}\right), \\
\padds &= \frac{\ln 2}{c}\left(1 + \frac{r(2 \ln 2 - 1)}{2 \ln 2 (1 - \ln 2)} + O(r^2)\right) + O\left(\frac{1}{c^2}\right).
\end{align}
\end{proposition}

\begin{IEEEproof}
Working out $a$, $a_0$ and $a_1$, and substituting them into $I(p) = p d(a_1 \| a) + (1 - p) d(a_0 \| a)$, we obtain
\begin{align}
I(p) &= p d(1 \| 1 - (1 - p)^c(1 - r)) \\
&+ (1 - p) d((1 - p)^{c-1} (1 - r) \| (1 - p)^c (1 - r)). 
\end{align}
For similar reasons as for the all-$1$ attack, for small values of $r$ the second term is $O(\frac{1}{c^2})$ while the first term is $\Theta(\frac{1}{c})$ and dominates the expression for large $c$. This means that for small $r$ we have
\begin{align}
I(p) &= -p \log_2(1 - (1 - p)^c (1 - r)) + O\left(\frac{1}{c^2}\right).
\end{align}
To find the maximum we take the derivative with respect to $p$ and set it equal to $0$ to obtain
\begin{align}
\ln(1 - (1 - p)^c (1 - r)) = -\frac{cp}{1 - p} \cdot \frac{(1 - p)^c (1 - r)}{1 - (1 - p)^c (1 - r)}.
\end{align}
For small $r$, the above expression is very close to the one we had for the all-$1$ attack, and again the optimal value of $p$ is close to $\frac{\ln 2}{c}$. Writing $s = (1 - p)^c (1 - r)$, so that $p = \frac{-1}{c} \ln(\frac{s}{1 - r}) + O(\frac{1}{c^2})$ and $1 - p = 1 - O(\frac{1}{c})$, the above expression reduces to
\begin{align}
\ln(1 - s) = \ln\left(\frac{s}{1 - r}\right) \cdot \frac{s}{1 - s} + O\left(\frac{1}{c}\right).
\end{align}
For small $r$, this means that $s \approx \frac{1}{2}$, so suppose $s = \frac{1}{2} (1 + \eps)$. Filling this in in the above equation, Tayloring around $\eps = 0$, and disregarding terms of the order $\eps^2, r^2, \eps r$, we get
\begin{align}
-\ln 2 - \eps = (-\ln 2 + r + \eps) (1 + 2\eps).
\end{align}
Rearranging the terms, this leads to
\begin{align}
\eps = -\frac{r}{2(1 - \ln 2)} + O(r^2).
\end{align}
Substituting $\eps$ into $s$ and solving for $p$, we get
\begin{align}
p &= -\frac{1}{c} \ln\left(\frac{1}{2} \cdot \frac{1 - \frac{r}{2(1 - \ln 2)}}{1 - r}\right) + O\left(\frac{1}{c^2}\right) \\
 &= \frac{\ln 2}{c} + \frac{r}{c} \cdot \frac{2 \ln 2 - 1}{2 - 2 \ln 2} + O\left(\frac{r^2}{c} + \frac{1}{c^2}\right),
\end{align}
and for the capacity we get
\begin{align}
I(p) &= -\frac{p}{\ln 2} \ln(1 - s) \\
 &= \left[-\frac{1}{c} + \frac{r}{c \ln 2} \cdot \frac{2 \ln 2 - 1}{2 - 2 \ln 2}\right] \left[- \ln 2 + \frac{r}{c} \cdot \frac{1}{2 - 2 \ln 2}\right] \\
 &= \frac{\ln 2}{c}\left(1 - \frac{r}{\ln 2} + O(r^2)\right) + O\left(\frac{1}{c^2}\right).
\end{align}
\end{IEEEproof}

For small values of $r$, one should therefore take $p$ to be slightly smaller than $\frac{1}{c} \ln 2$, and the capacity will be slightly lower than in the classical model. 


\subsubsection{Dilution noise}
\label{sec:cap-simple-dil}

Another commonly considered noisy group testing model is the dilution noise model~\cite{atia12, cheraghchi09, cheraghchi11, hwang76b, sejdinovic10}, where the probability of a positive test outcome depends on the number of defectives in the tested pool. More precisely, $\thdil$ is defined as follows:
\begin{align}
(\thdil)_z = \begin{cases} 0 & \text{if } z = 0; \\ 1 - r^z & \text{if } z > 0.\end{cases}
\end{align}
Again, for small $r$ this model is close to the traditional group testing model, so both the capacity and the optimal value of $p$ are close to the values of Proposition~\ref{thm:cap-simple-all}.

\begin{proposition} \label{thm:cap-simple-dil}
For the dilution noise model with parameter $r$, neglecting terms of the order $c^{-2}$ and $r^2$, the simple capacity and the corresponding optimal value of $p$ are:
\begin{align}
\caps(\thdil) &= \frac{\ln 2}{c} \left(1 + \frac{r \ln r}{2 \ln 2} - \frac{r(1 - \ln 2)}{2 \ln 2} + O(r^2 \ln r)\right) + O\left(\frac{1}{c^2}\right) \\
\pdils &= \frac{\ln 2}{c} \bigg(1 + \frac{r \ln r}{4 \ln 2} + \frac{r(-3 (\ln 2)^2 + 5 \ln 2 - 1)}{4 \ln 2(1 - \ln 2)} \\
 &+ O(r^2 \ln r)\bigg) + O\left(\frac{1}{c^2}\right).
\end{align}
\end{proposition}

\begin{IEEEproof}
For $a$, $a_0$ and $a_1$ we get
\begin{align}
a &= 1 - (1 - p + pr)^c, \\
a_0 &= 1 - (1 - p + pr)^{c-1}, \\
a_1 &= 1 - r(1 - p + pr)^{c-1},
\end{align}
so letting $s = (1 - p + pr)^c$, the mutual information satisfies
\begin{align}
I(p) &= p d\left(\frac{r s}{1 - p + pr} \| s\right) + (1 - p) d\left(\frac{s}{1 - p + pr} \| s\right). 
\end{align}
For small $r$, the second term is again small. So expanding the left term, knowing that $p = \Theta(\frac{1}{c})$, we obtain:
\begin{align}
I(p) &= \frac{p}{\ln 2} \left(r s \ln r + (1 - rs) \ln\left(\frac{1 - rs}{1 - s}\right)\right). 
\end{align}
Writing $p = \frac{\ln 2}{c} (1 + \eps)$, we can Taylor $s$ and $rs$ (disregarding terms of the order $r^2, r \eps^2, \eps^3, \frac{1}{c}$) to obtain
\begin{align}
s &= \frac{1}{2} \left(1 - \eps \ln 2 + r \ln 2 + \eps r \ln 2(1 - \ln 2) + \frac{\eps^2}{2} (\ln 2)^2\right).
\end{align}
This means that up to small order terms, we get $rs = \frac{1}{2}(r - \eps r \ln 2)$. Plugging these into the expression for $I(p)$, we eventually get
\begin{align}
&I(p) = \frac{\ln 2}{c} \Big(1 + r \left(\frac{\ln r - 1 + \ln 2}{2 \ln 2}\right) + \eps^2 \left(\ln 2 - 1\right) \\
 &\ + \eps r \left(\frac{\ln r(1 - \ln 2) - 3 (\ln 2)^2 + 5 \ln 2 - 1}{2 \ln 2}\right) + O(\dots)\Big).
\end{align}
This immediately leads to the given expression for the capacity by disregarding small terms, while differentiating with respect to $\eps$ and setting equal to $0$ leads to
\begin{align}
\eps &= \left(\frac{\ln r(1 - \ln 2) - 1 + 5 \ln 2 - 3 (\ln 2)^2}{4 \ln 2 (1 - \ln 2)}\right) r + O(r^2).
\end{align}
This leads to the given expression for $p$.
\end{IEEEproof}


\subsubsection{Threshold without gaps}
\label{sec:cap-simple-th1}

Besides accounting for possible mistakes in the test results (noisy group testing), models have also been considered to account for sensitivity in detecting positive items. In threshold group testing~\cite{ahlswede11, ahlswede11b, ahlswede13, chan13, cheraghchi13, damaschke06, lebedev10}, it is assumed that if the number of defectives $z$ in the tested pool is at most $l$ then the test comes back negative, and if $z$ is at least $u$ then the test result is always positive. For the case $u = l + 1$, which we will refer to as threshold group testing without a gap (where $g = u - l - 1$ is the gap size), this completely determines the model:
\begin{align}
(\ththr^{(u)})_z = \begin{cases} 0 & \text{if } z < u; \\ 1 & \text{if } z \geq u. \end{cases}
\end{align}
Although simple to state, even for small $u$ and $c$ finding the simple capacity and optimal choice of $p$ analytically seems very hard, if not impossible. We can intuitively see how the capacity will roughly behave though, since we know that:
\begin{itemize}
  \item The case $u = 1$ corresponds to $\ththr^{(u)} = \thall$, for which $p = \frac{\ln 2}{c}$ and $I \approx \frac{\ln 2}{c} \approx \frac{1.44}{c}$ are optimal.
  \item The case $u = \frac{c+1}{2}$ corresponds to $\ththr^{(u)} = \thmaj$, for which $p = \frac{1}{2}$ and $I = \frac{1}{\pi c \ln 2} \approx \frac{0.46}{c}$ are optimal.
\end{itemize}  
For values of $u$ between $1$ and $\frac{c}{2}$, we expect the capacity to decrease as $u$ increases, and the optimal value $p$ is expected to be close to $\frac{u}{c}$. 

Numerical evidence supports this intuition, as it shows that the capacity strictly decreases from $u = 1$ up to $u = \frac{c+1}{2}$, and that the optimal values of $p$ are almost evenly spaced for $u = 1$ up to $u = \frac{c}{2}$. The capacity quickly drops at small values of $u$, i.e., the gap between $\caps(\ththr^{(1)})$ and $\caps(\ththr^{(2)})$ is bigger than the gap between $\caps(\ththr^{(2)})$ and $\caps(\ththr^{(13)})$ for $c = 25$.



\subsubsection{Threshold with gaps}
\label{sec:cap-simple-th2}

An even harder case to deal with is threshold group testing with $g = u - l - 1 > 0$, which we will refer to as threshold group testing with a gap. If $u > l + 1$, then the model is not yet defined properly, as we do not know what $\theta_z$ is for $l + 1 \leq z \leq u - 1$. Different models were considered to capture the behavior of the outcome of the test results in these gaps, such as:~\cite{chan13}
\begin{itemize}
  \item The test outcome is uniformly random: 
\begin{align}
(\thcoi^{(l,u)})_z = \begin{cases} 0 & \text{if } z \leq l; \\ \frac{1}{2} & \text{if } l < z < u; \\ 1 & \text{if } z \geq u. \end{cases}
\end{align}
  \item The probability of a positive result increases linearly: 
\begin{align}
(\thint^{(l,u)})_z = \begin{cases} 0 & \text{if } z \leq l; \\ \frac{z - l}{u - l} & \text{if } l < z < u; \\ 1 & \text{if } z \geq u. \end{cases}
\end{align}  
  \item We simply do not know what the test outcome will be.
\end{itemize}
Note that $\thcoi^{(0,c)} = \thcoi$ and $\thint^{(0,c)} = \thint$, so these models can be seen as generalizations of the corresponding attacks in fingerprinting. Also note that $\thcoi^{(u-1, u)} = \thint^{(u-1, u)} = \ththr^{(u)}$.  

Regardless of the gap model, for arbitrary $l$ and $u$ these models all seem hard to analyze exactly. Using results obtained previously, we can however try to `interpolate' the results to get somewhat decent estimates. For instance, for the first model we can interpolate between the results for threshold group testing without a gap (Section~\ref{sec:cap-simple-th1}) and the coin-flip attack (Section~\ref{sec:cap-simple-coi}) to get upper and lower bounds on the simple capacity. For the second case, we can interpolate between threshold group testing without a gap (Section~\ref{sec:cap-simple-th1}) and the interleaving attack (Section~\ref{sec:cap-simple-int}) to get an idea how the capacity and the optimal value of $p$ scale. 

\begin{figure*}
\centering
\subfloat[][Simple capacity in the coin-flip gap model]{\includegraphics[width=0.45\textwidth]{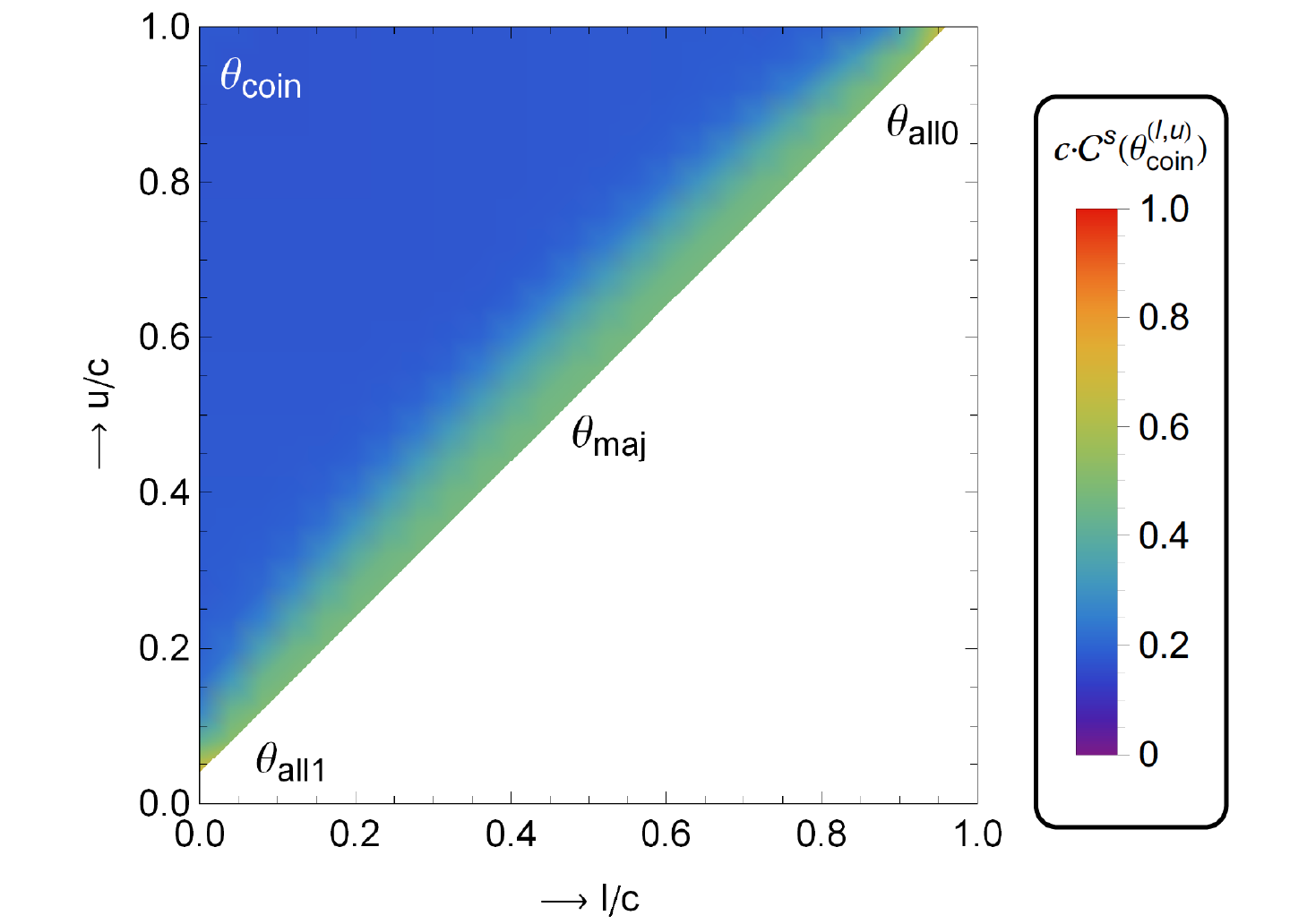}\label{fig:cap-simple-thres-coin}} \,
\subfloat[][Simple capacity in the interleaving gap model]{\includegraphics[width=0.45\textwidth]{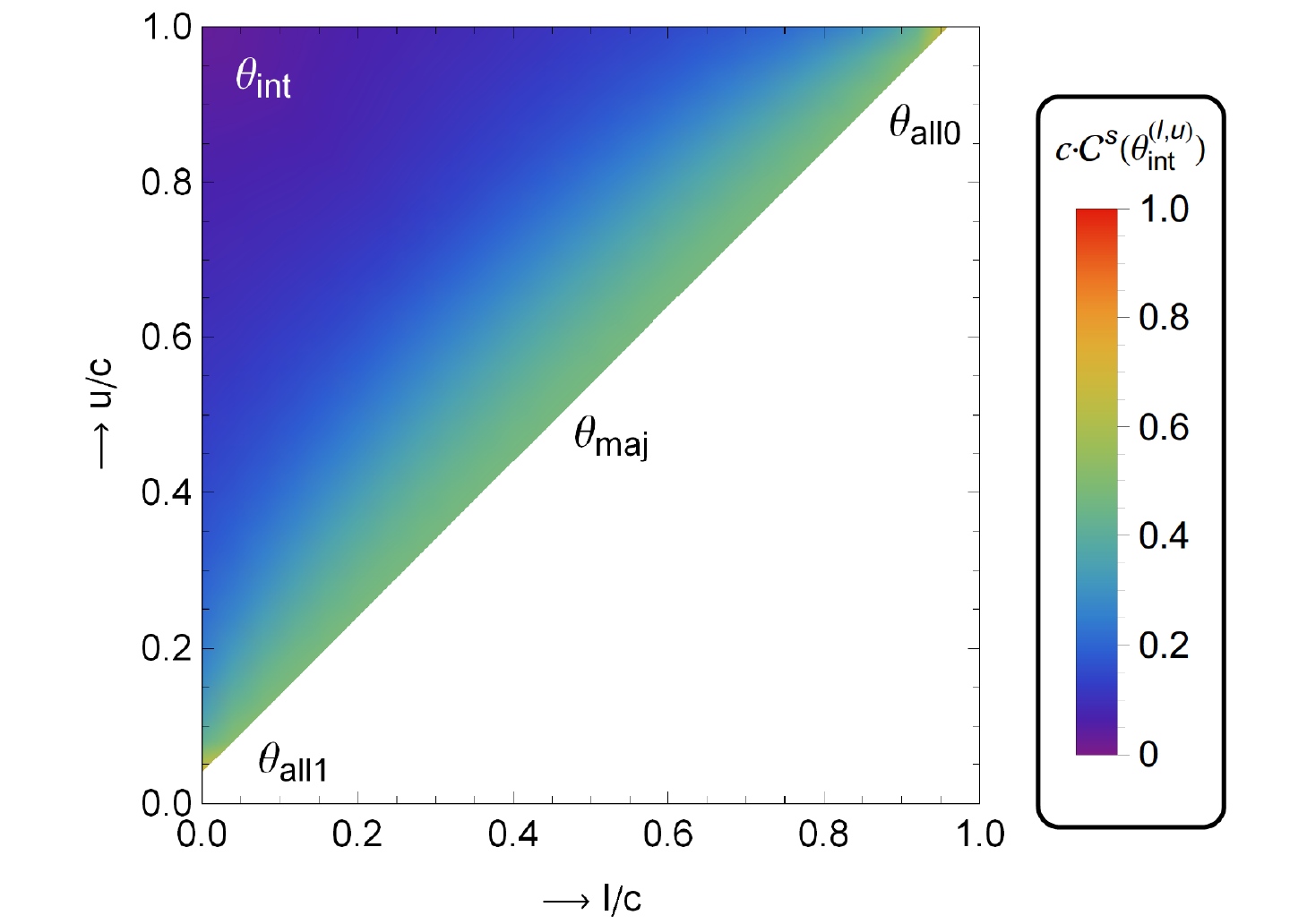}\label{fig:cap-simple-thres-int}}
\caption{The simple capacity for threshold group testing for different values of $l$ and $u$, when there is (a) a coin-flip gap or (b) an interleaving gap. The three corners correspond to the all-$1$, all-$0$ and coin-flip or interleaving attack, and the centers of the graphs correspond to the majority voting attack in fingerprinting. The capacity is maximal in the lower left and upper right corner, for which $c \cdot \capj(\vec{\theta}) \sim \ln 2 \approx 0.69$. \label{fig:cap-simple-thres}}
\end{figure*}

To verify this intuition, Figure~\ref{fig:cap-simple-thres} shows a density plot of the capacities (multiplied by $c$) for both the coin-flip gap model and the interleaving gap model. These plots are based on numerics for $c = 25$, but already show some trends. For instance, there are sharp peaks in the lower left and upper right corner; even when moving on the diagonal, the capacity quickly drops when leaving the corners. The capacities further take their maxima on and near the diagonal. In the coin-flip gap model, the capacity quickly converges to its minimum at $g = c$ as the gap size increases, while this takes longer for the interleaving gap model. Finally, from Sections~\ref{sec:cap-simple-int}, \ref{sec:cap-simple-all}, \ref{sec:cap-simple-maj}, and \ref{sec:cap-simple-coi}, we know exactly how the corners and center of each plot behave asymptotically, so we have a decent idea how the capacity scales for large $c$ and arbitrary values of $l$ and $u$.


\section{Joint capacities}
\label{sec:cap-joint}

Where a simple decoder bases its decision to accuse user $j$ only on the $j$th code word of $\mathcal{X}$ (and not on other code words), a joint decoder is allowed to use all information available to make a more informed decision. In particular, the whole code $\mathcal{X}$ may be taken into account. Huang and Moulin~\cite{huang09, huang09b, huang10, huang12} previously studied joint capacities as well, and showed that given a set of allowed collusion channels $\mathcal{P}^c$ (depending only on $\vec{\theta}$) and a set of allowed encoders $\mathcal{P}^e$, any fingerprinting rate below
\begin{align}
\capj(\mathcal{P}^e, \mathcal{P}^c) = \max_{f_P \in \mathcal{P}^e} \min_{\vec{\theta} \in \mathcal{P}^c} \expn_P I(p, \vec{\theta})
\end{align}
is achievable, where
\begin{align}
I(p, \vec{\theta}) = \frac{1}{c} I(X_1, \dots, X_c; Y | P = p)
\end{align}
is the mutual information between all colluder symbols $X_1, \dots, X_c$ and the pirate output $Y$ in one segment $i$. Note that from the assumption that $Y$ only depends on $X_1, \dots, X_c$ through $\vec{\theta}$, it follows that $I(X_1, \dots, X_c; Y | P = p) = I(Z; Y | P = p)$, where $Z = \sum_{i=1}^c X_i$. To study the payoff function $I(p, \vec{\theta}) = I(Z; Y|P = p)$, we will use the following identity~\cite[Equation~(59)]{huang12}:
\begin{align}
I(p, \vec{\theta}) &= \frac{1}{c} \left[h(a) - a_h\right] \\ 
\text{with } a_h &= \sum_{z = 0}^c \binom{c}{z} p^z (1 - p)^{c-z} h(\theta_z).
\end{align}
Here $h(\cdot)$ denotes the binary entropy function, defined by $h(\alpha) = -\alpha \log_2 \alpha - (1 - \alpha) \log_2 (1 - \alpha)$. Given $p$ and $\vec{\theta}$, this allows us to compute $I(p, \vec{\theta})$ explicitly. In the analysis of specific models $\vec{\theta}$, we will again commonly omit $\vec{\theta}$ as an argument of $I$ and write $I(p)$.

For obtaining the joint capacities for various models, we will extensively work with the binary entropy function. Again, this function can be quite ugly for arbitrary arguments $\alpha$, but in some cases we can somewhat simplify the expressions. For instance, for arguments close to $0$ or $\frac{1}{2}$ we have
\begin{align}
h(\gamma) &= \frac{\gamma (1 - \ln \gamma)}{\ln 2} - O(\gamma^2) = O(\gamma \ln \gamma), \\
h\left(\frac{1}{2} \pm \gamma\right) &= 1 - \frac{2\gamma^2}{\ln 2} - O(\gamma^4) = 1 - O(\gamma^2).
\end{align}
The most important properties to keep in mind are that $h(0) = h(1) = 0$ and $h$ takes its maximum at $\alpha = \frac{1}{2}$ with $h(\frac{1}{2}) = 1$. Using only these latter properties, we immediately get the following lemma regarding deterministic attacks, i.e., attacks satisfying $\vec{\theta} \in \{0,1\}^{c+1}$.

\begin{lemma} \label{lem:cap-joint-det}
For any deterministic attack $\vec{\theta}$ satisfying the marking assumption $\theta_0 = 0$ and $\theta_c = 1$, the joint capacity equals $\capj(\vec{\theta}) = \frac{1}{c}$, and $p$ is a maximizing value if it satisfies
\begin{align}
a = \sum_{z: \ \theta_z = 1} \binom{c}{z} p^z (1 - p)^{c-z} = \frac{1}{2} \, .
\end{align}
\end{lemma}

\begin{IEEEproof}
Since $\theta_z \in \{0,1\}$ for all $z$, we have $h(\theta_z) = 0$ for each $z$, so $a_h = 0$ and it thus follows that
\begin{align}
\capj(\vec{\theta}) = \max_p \frac{1}{c} \left[h(a) - a_h\right] = \frac{1}{c} \max_p h(a).
\end{align}
Since $a = a(p)$ is continuous in $p$, and $a(0) = 0$ and $a(1) = 1$ due to the marking assumption, from the intermediate value theorem it follows that there must be a value $p \in (0,1)$ for which $a(p) = \frac{1}{2}$. So we get
\begin{align}
\capj(\vec{\theta}) = \frac{1}{c} \max_p \left[h(a)\right] = \frac{1}{c} h\left(\frac{1}{2}\right) = \frac{1}{c} \, ,
\end{align}
and $p$ is a maximizing value iff $a(p) = \frac{1}{2}$.
\end{IEEEproof}

This lemma makes finding the joint capacities and the optimal values of $p$ very easy for several of the following models.


\subsection{Fingerprinting}
\label{sec:cap-joint-fp}


\subsubsection{Interleaving attack}
\label{sec:cap-joint-int}

We previously saw that the simple capacity for the interleaving attack is proportional to $\frac{1}{c^2}$. The exact asymptotics for the joint capacity were previously derived by Huang and Moulin as well, showing that for large $c$ the joint capacity is the same as the simple capacity.

\begin{proposition} \cite[Corollary 6]{huang12} \label{prop:cap-joint-int}
The joint capacity for the interleaving attack $\thint$ is:
\begin{align}
\capj(\thint) &= \frac{1}{2c^2 \ln 2} + O\left(\frac{1}{c^4}\right) \approx \frac{0.72}{c^2} \,.
\end{align}
\end{proposition}

Asymptotically, the interleaving attack is an ``equalizing strategy"~\cite{huang12}, guaranteeing that up to order terms $I(p)$ is the same for all $p \in (0,1)$.


\subsubsection{All-$1$ attack}
\label{sec:cap-joint-all}

Since the all-$1$ attack is a deterministic attack and satisfies the marking assumption, the capacity follows immediately from Lemma~\ref{lem:cap-joint-det}, and finding the optimal value of $p$ is straightforward.

\begin{proposition}
For the all-$1$ attack, the joint capacity and the maximizing value of $p$ are:
\begin{align}
\capj(\thall) &= \frac{1}{c} \, , \qquad \pallj = \frac{\ln 2}{c} + O\left(\frac{1}{c^2}\right).
\end{align}
\end{proposition}

To be precise, the optimal value for $p$ is $p = 1 - 2^{-1/c}$.


\subsubsection{Majority voting}
\label{sec:cap-joint-maj}

Lemma~\ref{lem:cap-joint-det} also applies to the majority voting attack $\thmaj$, and since this attack is symbol-symmetric the optimal value for $p$ is trivially $p = \frac{1}{2}$.

\begin{proposition}
For the majority voting attack, the joint capacity and the corresponding optimal value of $p$ are:
\begin{align}
\capj(\thmaj) &= \frac{1}{c} \, , \qquad \pmajj = \frac{1}{2} \, .
\end{align}
\end{proposition}

Note that the joint capacity for majority voting is equal to the joint capacity of the all-$1$ attack, while their simple capacities differ. Also note that again, the optimal value for $p$ is asymptotically the same as for the simple capacity.


\subsubsection{Minority voting}
\label{sec:cap-joint-min}

Since minority voting is also a deterministic and symbol-symmetric attack, the following result directly follows from Lemma~\ref{lem:cap-joint-det}.

\begin{proposition}
The joint capacity and a corresponding optimal value of $p$ for the minority voting attack are:
\begin{align}
\capj(\thmin) &= \frac{1}{c} \, , \qquad \pminj = \frac{1}{2} \, .
\end{align}
\end{proposition}

In fact, there are three values of $p$ that are asymptotically optimal, the other two being $p \approx \frac{\ln 2}{c}$ and $p \approx 1 - \frac{\ln 2}{c}$.


\subsubsection{Coin-flip attack}
\label{sec:cap-joint-coi}

Besides the interleaving attack, the only other non-trivial fingerprinting attack with respect to joint capacities is the coin-flip attack. This attack is not deterministic, so $a_h > 0$. Working out the details, we obtain the following result.

\begin{proposition} \label{prop:cap-joint-coi}
For the coin-flip attack, the joint capacity and the maximizing value of $p$ are:
\begin{align}
\capj(\thcoi) &= \frac{\log_2(5/4)}{c} + O\left(\frac{1}{c^2}\right) \approx \frac{0.32}{c} \, , \\
\pcoij &= \frac{\ln(5/3)}{c} + O\left(\frac{1}{c^2}\right) \approx \frac{0.51}{c} \, .
\end{align}
\end{proposition}

\begin{IEEEproof}
For $a_h$, note that $h(\theta_0) = h(\theta_c) = 0$ and $h(\theta_z) = 1$ otherwise, so $a_h = 1 - p^c - (1 - p)^c$. For $a$, recall from the proof of Proposition~\ref{thm:cap-simple-coi} that $a = \frac{1}{2}(1 - (1 - p)^c + p^c)$. Combining the above, we get
\begin{align}
I(p) &= \frac{1}{c}\left[h\left(\frac{1 - (1 - p)^c + p^c}{2}\right) - (1 - p^c - (1 - p)^c)\right].
\end{align}
Since the attack is symbol-symmetric, w.l.o.g.\ we may assume that $p \leq \frac{1}{2}$, in which case the terms $p^c$ are negligible for large $c$. Writing $t = 1 - (1 - p)^c$, we get 
\begin{align}
I(t) &= \frac{1}{c}\left[h\left(\frac{t}{2}\right) - t\right] + O\left(\frac{1}{c^2}\right).
\end{align}
This function has a maximum at $t = 1 - (1 - p)^c = \frac{2}{5}$, which leads to the given values of $\pcoij$ and $\capj(\thcoi)$.
\end{IEEEproof}


\subsubsection{Unknown attacks}
\label{sec:cap-joint-unk}

Finally, the case where the attack is not known (but is assumed to satisfy the marking assumption) was previously solved by Huang and Moulin, showing that again the interleaving attack is asymptotically optimal.

\begin{proposition} \cite[Corollary 7]{huang12}
The joint uninformed capacity is given by
\begin{align}
\capj(\pmark) &= \frac{1}{2 c^2 \ln 2} + O\left(\frac{1}{c^3}\right) \approx \frac{0.72}{c^2} \, ,
\end{align}
and the optimizing encoder $f_P$ and collusion channel $\vec{\theta}$ achieving this bound for large $c$ are the arcsine distribution $f_P^*$ and the interleaving attack $\thint$.
\end{proposition}

So while the joint capacities are asymptotically the same as the simple capacities for the interleaving attack and for the uninformed setting, for several other attacks the joint capacities are strictly higher than the simple capacities. 


\subsection{Group testing}
\label{sec:cap-joint-gt}


\subsubsection{Classical model}
\label{sec:cap-joint-ord}

Since the classical model is equivalent to the all-$1$ attack in group testing, the following result is immediate.

\begin{corollary}
For the classical group testing model, the joint capacity and the optimal value of $p$ are:
\begin{align}
\capj(\thall) &= \frac{1}{c} \, , \qquad \pallj = \frac{\ln 2}{c} + O\left(\frac{1}{c^2}\right).
\end{align}
\end{corollary}

This result was previously derived by Seb\H{o}~\cite[Theorem~2]{sebo85}, who also showed that $p = 1 - 2^{-1/c} \approx \frac{\ln 2}{c}$ is optimal. 


\subsubsection{Additive noise}
\label{sec:cap-joint-add}

The additive noise model described in Section~\ref{sec:cap-simple-add} was previously studied in the context of capacities in e.g.~\cite{atia12, cheraghchi11, sejdinovic10}. Cheraghchi et al.~\cite{cheraghchi11} showed that $\capj(\thadd) = O(\frac{(1 - r)^3}{c})$, while Atia and Saligrama~\cite{atia12} showed that $\capj(\thadd) = O(\frac{1 - r}{c})$. Looking closely at their proof, they show that one obtains a capacity of $I(p) \geq \frac{1 - r}{e c \ln 2} \approx \frac{1.88(1 - r)}{c}$ using $p = \frac{1}{c}$ for large $c$.\footnote{The authors of \cite{atia12} confirmed that the formula below \cite[(45)]{atia12} contains a typo: there should be an extra $e$ in the numerator of the code length $T$.} 

Below we improve upon these results, by (i) providing the exact leading constant on the capacity; (ii) showing exactly how the first order term (in $r$) scales for small $r$; and (iii) showing how $p$ scales in terms of $r$.

\begin{proposition}
For the additive noise model, the joint capacity and the corresponding optimal value of $p$ are:
\begin{align}
\capj(\thadd) &= \frac{1}{c}\left(1 - \tfrac{1}{2}h(r) + O(r^2)\right) + O\left(\frac{1}{c^2}\right), \\ 
\paddj &= \frac{\ln 2}{c} \left(1 - \frac{r(1 + \ln r)}{2 \ln 2} + O(r^2)\right) + O\left(\frac{1}{c^2}\right).
\end{align}
\end{proposition}

\begin{IEEEproof}
First, from the definition of $\thadd$ it follows that $a = 1 - (1 - p)^c (1 - r)$, $h(\theta_0) = h(1 - r)$ and $h(\theta_z) = 0$ for $z > 0$. So the mutual information satisfies
\begin{align}
I(p) &= \frac{1}{c}\left[h((1 - p)^c(1 - r)) - (1 - p)^c h(1 - r)\right].
\end{align}
Writing $s = (1 - p)^c (1 - r)$ this can be simplified to
\begin{align}
I(s) &= \frac{1}{c}\left[h(s) - \frac{h(r)s}{1 - r}\right].
\end{align}
We want to maximize $I$, so we take the derivative with respect to $s$, and set it equal to $0$ to obtain a condition for $s$, and hence for $p$:
\begin{align}
\log_2\left(\frac{s}{1 - s}\right) = -\frac{h(r)}{1 - r} \, .
\end{align}
For small $r$, the right hand side goes to $0$, which implies that $s$ is close to $\frac{1}{2}$. So assuming $r$ is small, we let $s = \frac{1}{2} (1 + \eps)$ and obtain the following Taylor expansion for the left hand side:
\begin{align}
\log_2\left(\frac{1 + \eps}{1 - \eps}\right) = \log_2\left(1 + 2 \eps + O(\eps^2)\right) = \frac{2\eps}{\ln 2} + O(\eps^2).
\end{align}
This means that for small $r$, the optimal choice for $\eps$ is
\begin{align}
\eps 
 &= -\frac{h(r) \ln 2}{2} + O(r^2).
\end{align}
So for $s$ we obtain
\begin{align}
s = (1 - p)^c (1 - r) = \frac{1}{2}\left(1 - \tfrac{1}{2} h(r) \ln 2 + O(r^2)\right).
\end{align}
Substituting $s$ into $I(s)$, and solving $s$ for $p$, we obtain the given values for $\paddj$ and $\capj(\thadd)$.
\end{IEEEproof}

Note that this means that any valid group testing algorithm asymptotically requires at least the following number of tests:
\begin{align}
\ell &\geq \frac{c \log_2 n}{1 - \frac{1}{2} h(r) + O(r^2)} \left(1 + O\left(\frac{1}{c}\right)\right).
\end{align}
Since $r = o(h(r))$ for small $r$, this shows that the result of~\cite{atia12} is slightly off; due to their suboptimal choice of $p$, they obtained a code length which scales ``better" in $r$, but has a higher leading constant and thus converges to the wrong limit.


\subsubsection{Dilution noise}
\label{sec:cap-joint-dil}

The dilution noise model, as described in Section~\ref{sec:cap-simple-dil}, was previously studied in the context of lower bounds by Atia and Saligrama~\cite{atia12}. In terms of capacities, they showed that for large $c$, one has $\capj(\thdil) = O(\frac{(1 - r)^2}{c})$. Again, they were not interested in leading constants, so they fixed $p$ to the suboptimal choice $p = \frac{1}{c}$. We improve upon their result by finding the leading constant explicitly, and proving how $\pdilj$ and $\capj(\thdil)$ scale in terms of $r$.

\begin{proposition}
For the dilution noise model with parameter $r$, the joint capacity and the corresponding maximizing value of $p$ are:
\begin{align}
\capj(\thdil) &= \frac{1}{c} \left(1 - \frac{\ln 2}{2} h(r) + O(r^2)\right) + O\left(\frac{1}{c^2}\right), \\
\pdilj 
 &= \frac{\ln 2}{c} \left(1 + r - \frac{1 - \ln 2}{2} h(r) + O(r^2)\right) + O\left(\frac{1}{c^2}\right).
\end{align}
\end{proposition}

\begin{IEEEproof}
For this attack, we have $\theta_z = 1 - r^z$. Let us first look at $h(a)$:
\begin{align}
h(a) &= h\left(\sum_{z=0}^c \binom{c}{z} p^z (1 - p)^{c-z} (1 - r^z)\right) \\
 &= h(1 - (1 - p + pr)^c).
\end{align}
Next, consider $a_h$:
\begin{align}
a_h &= \sum_{z = 1}^c \binom{c}{z} p^z (1 - p)^{c-z} h(1 - r^z).
\end{align}
For small $r$, the only significant contribution to the sum comes from the term with $z = 1$:
\begin{align}
a_h &= c p (1 - p)^{c-1} h(r) + O(r^2).
\end{align}
The optimal value of $p$ again lies close to $\frac{\ln 2}{c}$; in particular, the value is mostly determined by the term $h((1 - p + pr)^c)$, which has a maximum at $(1 - p + pr)^c = \frac{1}{2}$. Writing $(1 - p + pr)^c = \frac{1}{2}(1 + \eps)$, we have
\begin{align}
p &= \frac{1}{c} \left(\ln 2 + r \ln 2 - \eps - r\eps + \frac{\eps^2}{2} + O(r^2, \eps^2 r, \eps^3)\right), \\
(1 - p)^c &= \frac{1}{2} \left(1 - r \ln 2 + \eps + r \eps - \frac{\eps^2}{2} + O(r^2, \eps^2 r, \eps^3)\right).
\end{align}
This means that $I(p) = I(\eps)$ satisfies (neglecting terms of the order $r^2, \eps^2 r, \eps^3, c^{-1}$)
\begin{align}
I(\eps) \sim 1 - \frac{1}{2} h(r) \ln 2 + \frac{1}{2} \eps h(r) (1 - \ln 2) - \frac{\eps^2}{2 \ln 2}.
\end{align}
Taking the derivative with respect to $\eps$ and setting it equal to $0$, we obtain
\begin{align}
\eps = \frac{1}{2} h(r) \ln 2 (1 - \ln 2) + O(r^2).
\end{align}
Substituting this value for $\eps$ in the expressions for $p$ and $I$, we get the results.
\end{IEEEproof}

For the resulting lower bound on the code length $\ell$, one thus obtains
\begin{align}
\ell \sim \frac{c \log_2 n}{1 - \frac{1}{2} h(r) \ln 2 + O(r^2)} \, .
\end{align}
So also in the dilution noise model, the first order term in the denominator scales as $h(r)$ rather than $r$, as one might suspect from the results of~\cite{atia12}.


\subsubsection{Threshold without gaps}
\label{sec:cap-joint-th1}

For threshold group testing with $u = l + 1$ (as described in Section~\ref{sec:cap-simple-th1}) we now consider two different cases for $u$: $u = \Theta(c)$ and $u = o(c)$. In both cases, the capacity follows directly from Lemma~\ref{lem:cap-joint-det}, but we can obtain slightly more accurate asymptotics for $p$ in the second case. The first case is sometimes referred to in the literature as majority group testing~\cite{ahlswede11, ahlswede11b, ahlswede13}.

\begin{proposition} \label{lem:cap-joint-thres-without}
For the threshold group testing model with $u = \ell + 1$, the joint capacity is $\frac{1}{c}$, and the corresponding maximizing value of $p$ is:
\begin{align}
u = \Theta(c): \quad \pthrj[\ththr^{(u)}] &= \frac{1}{c} \left(u + \gamma\right) 
\qquad (|\gamma| \leq 1) \\
u = o(c): \quad \pthrj[\ththr^{(u)}] &= \frac{1}{c} \left(u - \frac{1}{3} + O\left(\frac{1}{u}\right)\right). 
\end{align}
\end{proposition}

\begin{IEEEproof}
From Lemma~\ref{lem:cap-joint-det} it follows that the capacity is $\frac{1}{c}$ and that the optimal value of $p$ satisfies $a = \frac{1}{2}$. Writing out $a$, we have
\begin{align}
a &= \sum_{z = 0}^{u-1} \binom{c}{z} p^z (1 - p)^{c-z} = \frac{1}{2}.
\end{align}
The fact that $a = \frac{1}{2}$ roughly means that $u$ is the median of the binomial distribution with $c$ trials and probability of success $p$. Since the median of a binomial distribution is one of the two integers closest to $cp$, it follows that $|u - cp| \leq 1$ leading to the result for the case $u = \Theta(c)$. 

For the case $u = o(c)$, note that $p = O(\frac{1}{c})$, so $(1 - p)^z = 1 - O(p)$ for $z < u$. So we can expand $a$ around $c = \infty$ as:
\begin{align}
a &= (1 - p)^c \sum_{z = 0}^{u - 1} \binom{c}{z} p^z + O\left(\frac{1}{c}\right).
\end{align}
Since the solution is in the range $p = \Theta(\frac{1}{c})$, let us write $p = \frac{\alpha}{c}$ for some constant $\alpha$. A Taylor expansion around $c = \infty$ of the binomial coefficients then gives us
\begin{align}
a &= e^{-\alpha} \sum_{z = 0}^{u - 1} \frac{\alpha^z}{z!} + O\left(\frac{1}{c}\right).
\end{align}
The condition that $a = \frac{1}{2}$ means that asymptotically, $u - 1$ is the median of the Poisson distribution with parameter $\lambda = \alpha$. Using results about the median of the Poisson distribution~\cite{choi94}, we obtain
\begin{align}
\alpha = u - \frac{1}{3} + O\left(\frac{1}{u}\right).
\end{align}
Substituting this back into $p$, we get the result.
\end{IEEEproof}

Note that for $u = 1$ and $c \to \infty$, the above approximation says $p \approx \frac{0.67}{c}$, when in reality the optimum is at $p \sim \frac{\ln 2}{c} \approx \frac{0.69}{c}$, showing that already for small values of $u$ the term $u - \frac{1}{3}$ is quite accurate.


\subsubsection{Threshold with gaps}
\label{sec:cap-joint-th2}

\begin{figure*}
\centering
\subfloat[][Joint capacity in the coin-flip gap model]{\includegraphics[width=0.45\textwidth]{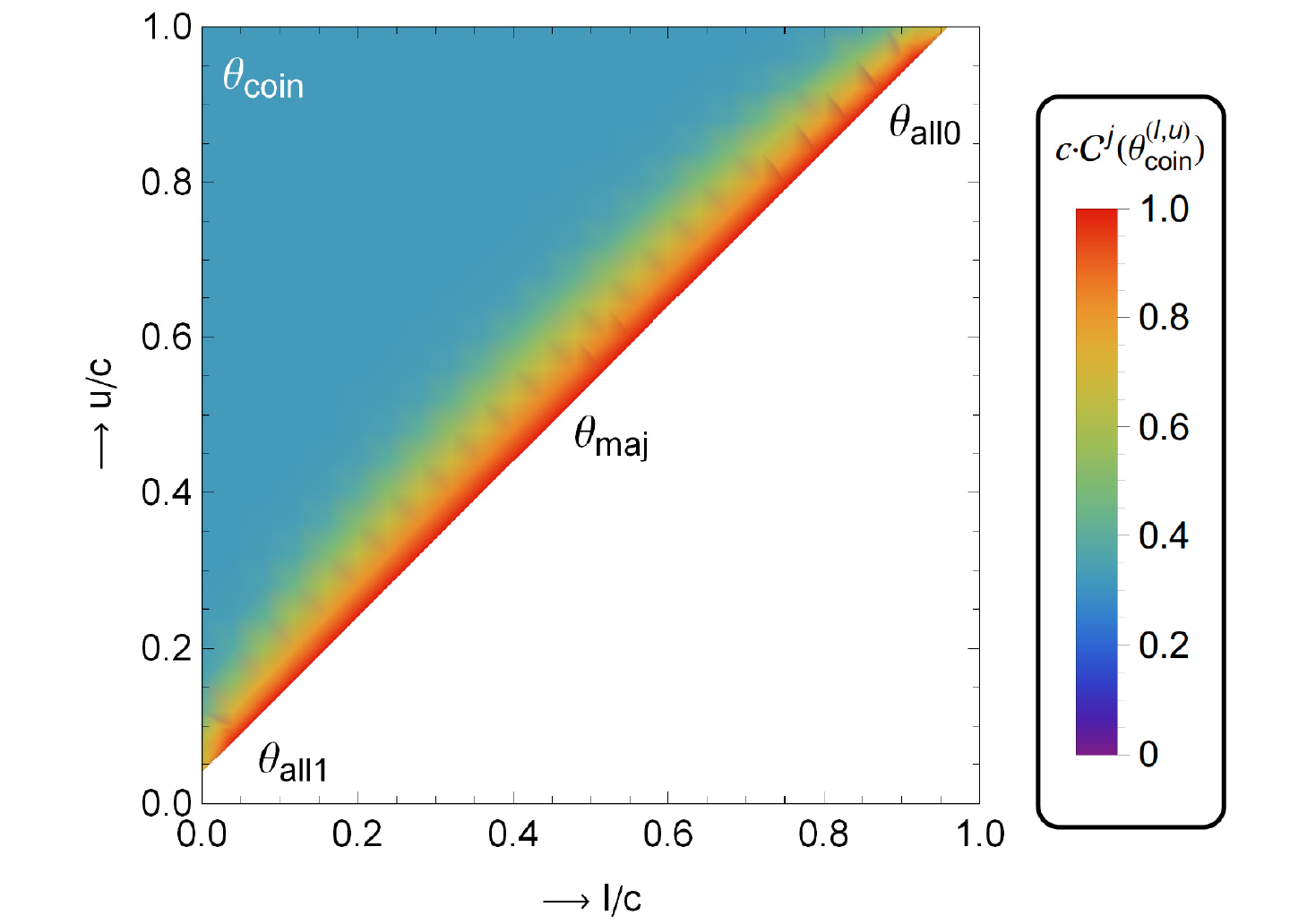}\label{fig:cap-joint-thres-coin}} \,
\subfloat[][Joint capacity in the interleaving gap model]{\includegraphics[width=0.45\textwidth]{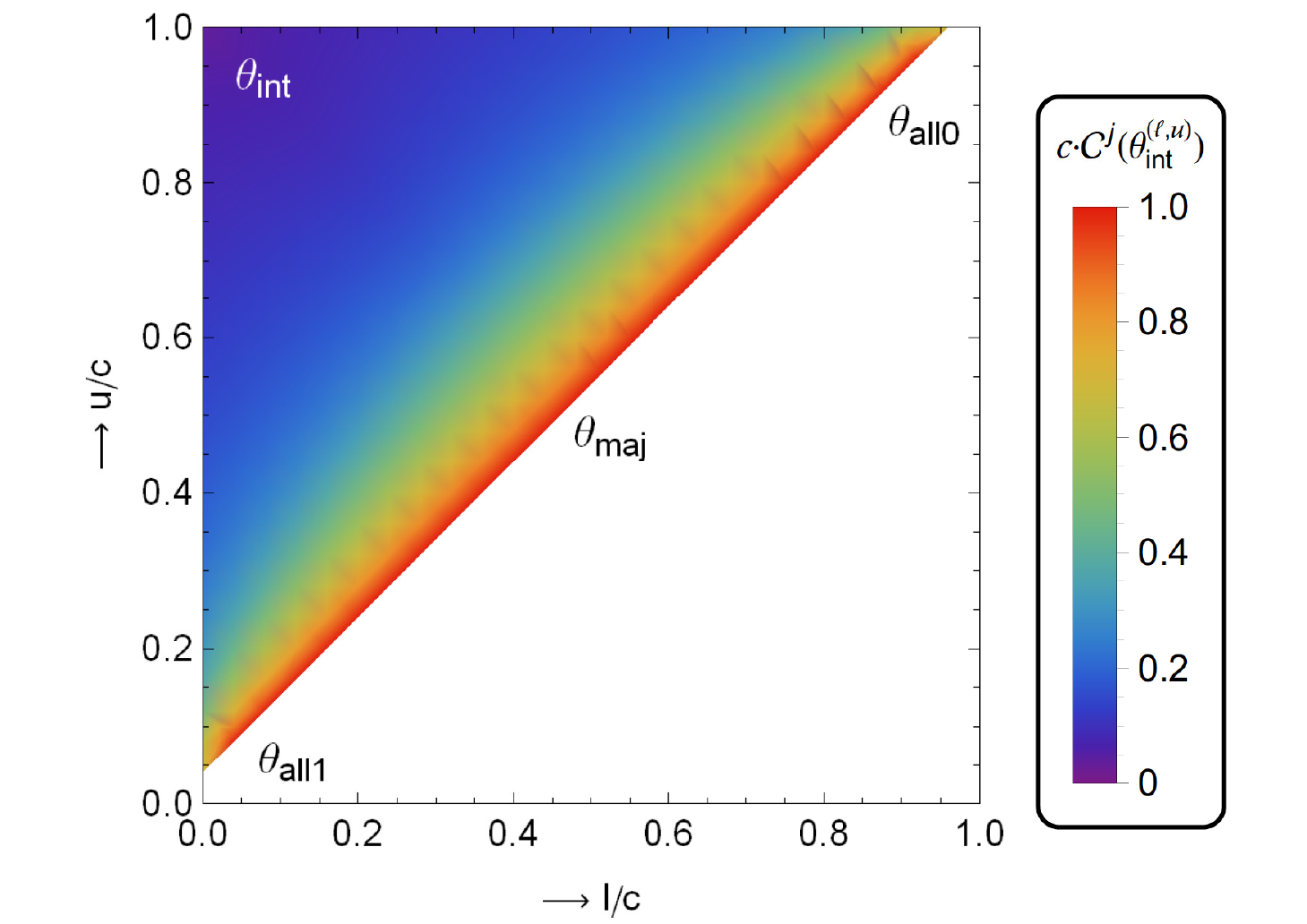}\label{fig:cap-joint-thres-int}}
\caption{The joint capacity for threshold group testing with different values of $l$ and $u$, when there is (a) a coin-flip gap or (b) an interleaving gap. The three corners correspond to the all-$1$, all-$0$ and coin-flip (a) or interleaving (b) attack, and the center of the graphs corresponds to the majority voting attack in fingerprinting. The capacity is maximal on the diagonal, for which $c \cdot \capj(\vec{\theta}) = 1$. \label{fig:cap-joint-thres}}
\end{figure*}

For threshold group testing with gaps, let us again consider the two models described in Section~\ref{sec:cap-simple-th2}: the coin-flip gap model and the interleaving gap model. For both models, we can again interpolate between results obtained earlier in this section to obtain estimates for $\capj(\thcoi^{(l, u)})$ and $\capj(\thint^{(l, u)})$ for various $l$ and $u$, and verify our intuition numerically (see Figure~\ref{fig:cap-joint-thres}). In both plots, from Proposition~\ref{lem:cap-joint-thres-without} it follows that the diagonals have value $c \cdot \capj(\vec{\theta}) = 1$, while the upper left corner in Figure~\ref{fig:cap-joint-thres-coin} converges to $\log_2(5/4) \approx 0.32$ (Proposition~\ref{prop:cap-joint-coi}) and the upper left corner of Figure~\ref{fig:cap-joint-thres-int} converges to $0$ (Proposition~\ref{prop:cap-joint-int}). In the left graph, even for small gaps we see that the capacity quickly decreases and approaches the coin-flip capacity. In the right graph, we see that the capacity decreases more gradually as the gap size increases.

%
\section{Discussion}
\label{sec:discussion}

Building upon previous work of Huang and Moulin and working our way through the resulting expressions for the capacities, we have derived explicit asymptotics for both the simple and joint capacities for various fingerprinting and group testing models. In the end the results from fingerprinting turned out to be useful in threshold group testing as well, for understanding the numerics of Figures~\ref{fig:cap-simple-thres} and \ref{fig:cap-joint-thres} and estimating the capacities for various threshold group testing models.

One important result with respect to group testing is that the simple capacity in the traditional model is asymptotically a factor $\log_2(e)$ lower than the joint capacity. While the joint capacity was well known, to the best of our knowledge the simple capacity had not yet been derived before. This result shows that efficient (simple) group testing algorithms will never be able to achieve the code lengths of optimal joint decoders, and that various existing methods (e.g.~\cite{chan11, chan12}) are suboptimal, even for simple decoding. The related paper~\cite{laarhoven14decoders} explicitly shows how the bounds on the code lengths of simple decoders can be attained with log-likelihood decoders.

Comparing the simple and joint capacities, another result worth mentioning is that except for in the cases previously analyzed by Huang and Moulin, there is always a gap between the simple and joint capacities. In fingerprinting, this means that if the pirates use a suboptimal attack, joint decoders are asymptotically significantly better than simple decoders. In terms of group testing, this means that in almost all models, simple decoders are strictly worse than joint decoders. So although joint decoders are generally slower, the benefits of joint decoding (a much shorter code length) may outweigh the costs of a higher decoding complexity.

%
\section{Open problems}
\label{sec:openproblems}

Let us finish by mentioning some open problems which are left for future work.


\subsection{Dynamic fingerprinting and adaptive group testing}

While this paper considered only static fingerprinting and non-adaptive group testing, in some settings the feedback $Y$ may be obtained in real-time. For instance, in pay-tv pirates may try to duplicate a fingerprinted broadcast, while in group testing it may sometimes be possible to do group tests sequentially. These dynamic or adaptive games have received considerable attention as well \cite{aldridge12, baldassini13, debonis05, fiat01, laarhoven12wifs, laarhoven13tit, laarhoven13wifs} but little is known about the capacities of these games. Are the dynamic/adaptive capacities strictly higher than the static/non-adaptive capacities in the probabilistic model considered in this paper?  


\subsection{Tuple decoders and tuple capacities}

Recall that simple decoders base their decisions only on individual code words, while joint decoders base their decisions on the entire code $\mathcal{X}$. The extra information used by joint decoders generally causes the joint capacity to be higher than the simple capacity, but the complexity of decoding may be higher as well. A possible way to obtain a trade-off between the code length and the time complexity would be \textit{tuple decoding}: basing the decision to accuse a user $j$ only on tuples of size at most $t$. This could be seen as a generalization of simple and joint decoding, since those models correspond to $t = 1$ and $t = c$ respectively. Such decoders were previously considered in e.g.~\cite{amiri09, meerwald12, moulin08, oosterwijk14}, and an obvious question is: can we somehow quantize this trade-off between the time complexity and the code length? And can we formally derive capacities for this tuple decoding model?


\subsection{Non-binary codes in fingerprinting}

A common generalization in fingerprinting is to assume that symbols come from an alphabet of size $q \geq 2$, rather than assuming that the code $\mathcal{X}$ is a binary code ($q = 2$). This generalization was considered in e.g.~\cite{boesten11, huang12b, oosterwijk13b, oosterwijk14, skoric08}. In the uninformed fingerprinting game, the capacity decreases linearly with $q$~\cite{boesten11, huang12b}, so there may be significant benefits going from a binary to a $q$-ary alphabet. For the models considered in this paper, for which the capacity is only linear in $c$, it is easy to see that the capacity cannot increase linearly with $q$. Some basic numerics seem to indicate that the capacity increases with a factor $\log q$, but a more detailed analysis is required.


\subsection{Universal encoding in fingerprinting}

\begin{figure*}[!ht]
\centering
\subfloat[][Simple capacity in the universal encoding model]{\includegraphics[width=0.45\textwidth]{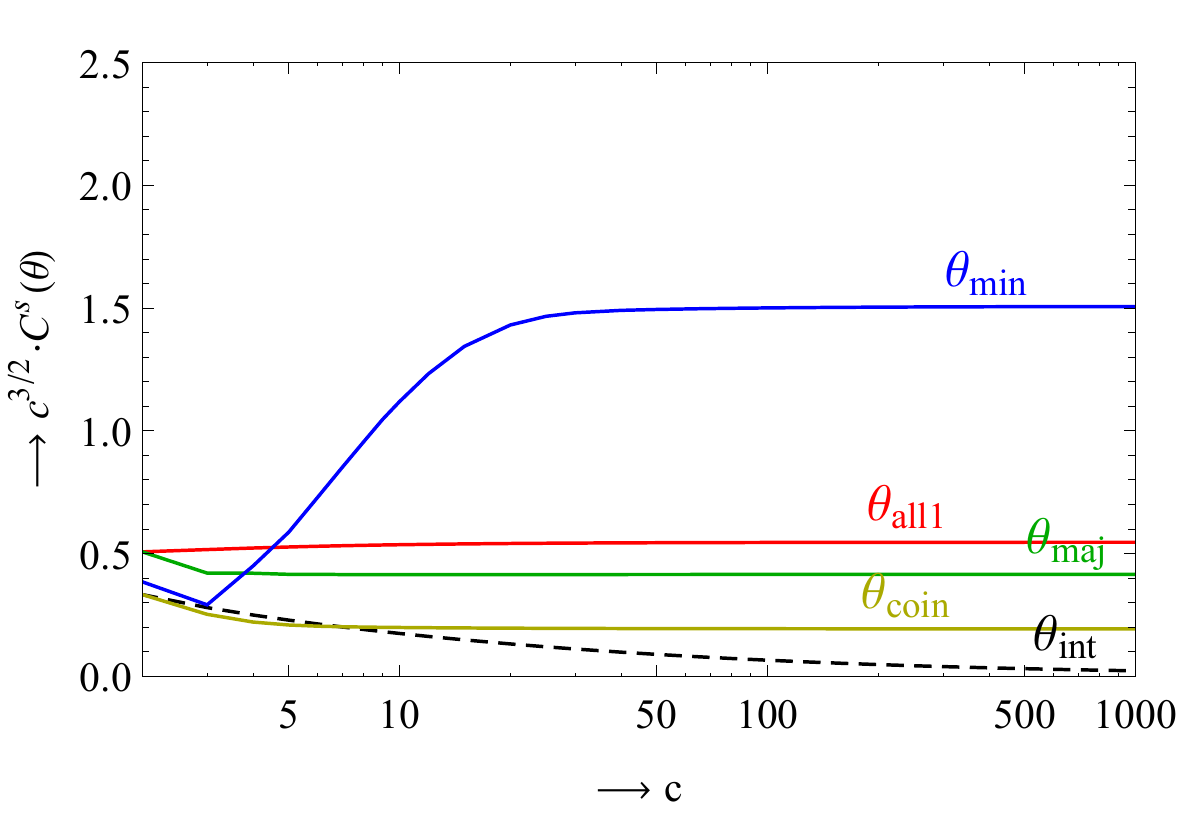}\label{fig:cap-simple-part}} \,
\subfloat[][Joint capacity in the universal encoding model]{\includegraphics[width=0.45\textwidth]{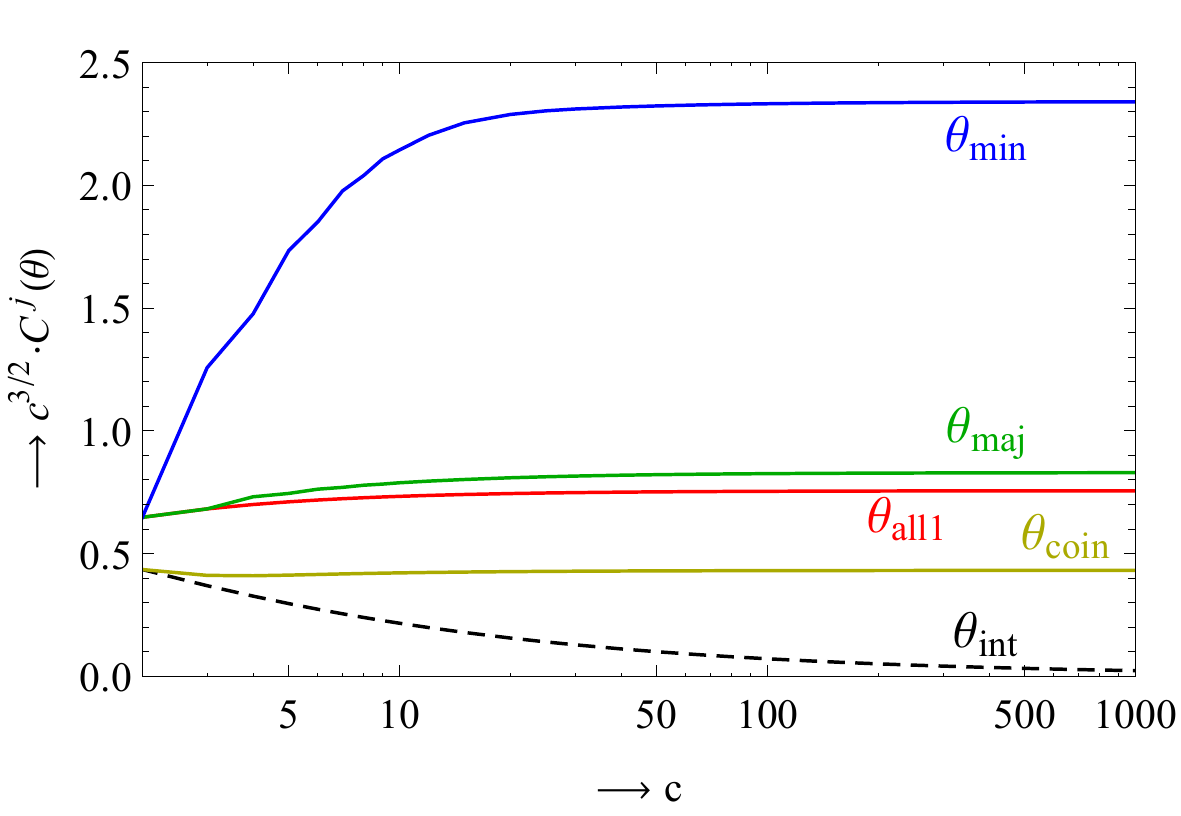}\label{fig:cap-joint-part}}
\caption{The simple (a) and joint (b) universal encoding capacities (multiplied by $c^{3/2}$) as a function of $c$, for different pirate attacks. Except for the interleaving attack, for which the capacity scales as $c^{-2}$ (the dashed line), these capacities all seem to scale as $c^{-3/2}$. \label{fig:cap-part}}
\end{figure*}

Finally, instead of assuming that the pirate attack is known in advance, in fingerprinting it is more often assumed that the encoding is done for arbitrary attacks, and that only the decoding step may be tuned to fit the pirate attack~\cite{charpentier09, furon09b, meerwald11b, oosterwijk13b}. Since the asymptotically optimal universal encoding strategy is to use the arcsine distribution $f_P^*$ for generating biases $p$, one could try deriving the capacities for the various fingerprinting attacks in case the distribution $f_P$ is fixed in advance as $f_P \equiv f_P^*$. Previous results~\cite{oosterwijk13b} showed that the capacities probably scale as $c^{-3/2}$, and numerics of the associated capacities (Figure~\ref{fig:cap-part}) seem to verify this. Obtaining exact expressions for the simple and joint capacities under `universal encoding' is left for future work.

\section*{Acknowledgments} 

The author is grateful to Benne de Weger for his help with some of the proofs in this paper, and for his comments on drafts of this manuscript that helped improve the paper. The author would further like to thank Jeroen Doumen, Teddy Furon, Jan-Jaap Oosterwijk, and Boris \v{S}kori\'{c} for their valuable comments and suggestions.



\begin{thebibliography}{99}

\bibitem{ahlswede11}
R.~Ahlswede, C.~Deppe, and V.~S.~Lebedev, ``Bounds for Threshold and Majority Group Testing," \emph{IEEE International Symposium on Information Theory (ISIT)}, pp.~69--73, 2011.

\bibitem{ahlswede11b}
R.~Ahlswede, C.~Deppe, and V.~S.~Lebedev, ``Majority Group Testing with Density Tests," \emph{IEEE International Symposium on Information Theory (ISIT)}, pp.~326--330, 2011.

\bibitem{ahlswede13}
R.~Ahlswede, C.~Deppe, and V.~S.~Lebedev, ``Threshold and Majority Group Testing," \emph{Information Theory, Combinatorics, and Search Theory}, LNCS vol.~7777, pp.~488--508, 2013.

\bibitem{aldridge12}
M.~Aldridge, ``Adaptive Group Testing as Channel Coding with Feedback," \emph{IEEE International Symposium on Information Theory (ISIT)}, pp.~1832--1836, 2012.


\bibitem{amiri09}
E.~Amiri and G.~Tardos, ``High Rate Fingerprinting Codes and the Fingerprinting Capacity," \emph{20th ACM-SIAM Symposium on Discrete Algorithms (SODA)}, pp.~336--345, 2009.

\bibitem{atia09}
G.~K.~Atia and V.~Saligrama, ``Noisy Group Testing: An Information Theoretic Perspective," \emph{47th Annual Allerton Conference on Communication, Control, and Computing (Allerton)}, pp.~355--362, 2009.

\bibitem{atia12}
G.~K.~Atia and V.~Saligrama, ``Boolean Compressed Sensing and Noisy Group Testing," \emph{IEEE Transactions on Information Theory}, vol.~58, no.~3, pp.~1880--1901, 2012.

\bibitem{baldassini13}
L.~Baldassini, O.~Johnson, and M.~Aldridge, ``The Capacity of Adaptive Group Testing," \emph{arXiv}, 2013.

\bibitem{berchtold12}
W.~Berchtold and M.~Sch\"{a}fer, ``Performance and Code Length Optimization of Joint Decoding Tardos Fingerprinting," \emph{ACM Symposium on Multimedia and Security (MMSec)}, pp.~27--32, 2012.

\bibitem{boesten11}
D.~Boesten and B.~\v{S}kori\'{c}, ``Asymptotic Fingerprinting Capacity for Non-Binary Alphabets," \emph{13th Conference on Information Hiding (IH)}, pp.~1--13, 2011.

\bibitem{boneh98}
D.~Boneh and J.~Shaw, ``Collusion-Secure Fingerprinting for Digital Data," \emph{IEEE Transactions on Information Theory}, vol.~44, no.~5, pp.~1897--1905, 1998.

\bibitem{chan11}
C.-L.~Chan, P.~H.~Che, S.~Jaggi, and V.~Saligrama, ``Non-adaptive probabilistic group testing with noisy measurements: Near-optimal bounds with efficient algorithms," \emph{49th Allerton Conference on Communication, Control, and Computing}, pp.~1832--1839, 2011.

\bibitem{chan12}
C.-L.~Chan, S.~Jaggi, V.~Saligrama, and S.~Agnihotri, ``Non-Adaptive Group Testing: Explicit Bounds and Novel Algorithms," \emph{IEEE International Symposium on Information Theory (ISIT)}, pp.~1837--1841, 2012.

\bibitem{chan13}
C.-L.~Chan, S.~Cai, M.~Bakshi, S.~Jaggi, and V.~Saligrama, ``Near-Optimal Stochastic Threshold Group Testing," \emph{arXiv}, 2013.

\bibitem{charpentier09}
A.~Charpentier, F.~Xie, C.~Fontaine, and T.~Furon, ``Expectation Maximization Decoding of Tardos Probabilistic Fingerprinting	Code," \emph{SPIE Proceedings / Media Forensics and Security}, vol.~7254, 2009.

\bibitem{chen09}
H.-B.~Chen and H.-L.~Fu, ``Nonadaptive Algorithms for Threshold Group Testing," \emph{Discrete Applied Mathematics}, vol.~157, no.~7, pp.~1581--1585, 2009.

\bibitem{cheraghchi09}
M.~Cheraghchi, A.~Hormati, A.~Karbasi, and M.~Vetterli, ``Compressed Sensing with Probabilistic Measurements: A Group Testing Solution," \emph{47th Allerton Conference on Communication, Control, and Computing (Allerton)}, pp.~30--35, 2009.


\bibitem{cheraghchi11}
M.~Cheraghchi, A.~Hormati, A.~Karbasi, and M.~Vetterli, ``Group Testing with Probabilistic Tests: Theory, Design and Application," \emph{IEEE Transactions on Information Theory}, vol.~57, no.~10, pp.~7057--7067, 2011.

\bibitem{cheraghchi13}
M.~Cheraghchi, ``Improved Constructions for Non-Adaptive Threshold Group Testing," \emph{Algorithmica}, vol.~67, no.~3, pp.~384--417, 2013.

\bibitem{choi94}
K.~P.~Choi, ``On the Medians of Gamma Distributions and an Equation of Ramanujan," \emph{Proceedings of the American Mathematical Society}, vol.~121, no.~1, pp.~245--251, 1994.

\bibitem{cover06}
T.~M.~Cover and J.~A.~Thomas, \emph{Elements of Information Theory (2nd Edition)}, Wiley Press, 2006.

\bibitem{damaschke06}
P.~Damaschke, ``Threshold Group Testing," \emph{General Theory of Information Transfer and Combinatorics}, LNCS vol.~4123, pp.~707--718, 2006.

\bibitem{debonis05}
A.~De~Bonis, L.~Gasieniec, and U.~Vaccaro, ``Optimal Two-Stage Algorithms for Group Testing Problems," \emph{SIAM Journal on Computing}, vol.~34, no.~5, pp.~1253--1270, 2005.

\bibitem{dorfman43}
R.~Dorfman, ``The Detection of Defective Members of Large Populations," \emph{The Annals of Mathematical Statistics}, vol.~14, no.~4, pp.~436--440, 1943.


\bibitem{dyachkov82}
A.~G.~D'yachkov and V.~V.~Rykov, ``Bounds on the length of disjunctive codes," \emph{Problemy Peredachi Informatsii}, vol.~18, no.~3, pp.~7--13, 1982.

\bibitem{dyachkov89}
A.~G.~D'yachkov, V.~V.~Rykov, and A.~M.~Rashad, ``Superimposed distance codes," \emph{Problems of Control and Information Theory}, vol.~18, no.~4, pp.~237--250, 1989.


\bibitem{fiat01}
A.~Fiat and T.~Tassa, ``Dynamic Traitor Tracing," \emph{Journal of Cryptology}, vol.~14, no.~3, pp.~211--223, 2001.

\bibitem{furon09b}
T.~Furon and L.~P\'{e}rez-Freire, ``EM Decoding of Tardos Traitor Tracing Codes," \emph{ACM Symposium on Multimedia and Security (MM\&Sec)}, pp.~99--106, 2009.

\bibitem{huang09}
Y.-W.~Huang and P.~Moulin, ``Capacity-Achieving Fingerprint Decoding," \emph{IEEE Workshop on Information Forensics and Security (WIFS)}, pp.~51--55, 2009.

\bibitem{huang09b}
Y.-W.~Huang and P.~Moulin, ``Saddle-Point Solution of the Fingerprinting Capacity Game under the Marking Assumption," \emph{IEEE International Symposium on Information Theory (ISIT)}, pp.~2256--2260, 2009.

\bibitem{huang10}
Y.-W.~Huang and P.~Moulin, ``Maximin Optimality of the Arcsine Fingerprinting Distribution and the Interleaving Attack for Large Coalitions," \emph{IEEE Workshop on Information Forensics and Security (WIFS)}, pp.~1--6, 2010.

\bibitem{huang12}
Y.-W.~Huang and P.~Moulin, ``On the Saddle-Point Solution and the Large-Coalition Asymptotics of Fingerprinting Games," \emph{IEEE Transactions on Information Forensics and Security}, vol.~7, no.~1, pp.~160--175, 2012.

\bibitem{huang12b}
Y.-W.~Huang and P.~Moulin, ``On Fingerprinting Capacity Games for Arbitrary Alphabets and Their Asymptotics," \emph{IEEE International Symposium on Information Theory (ISIT)}, pp.~2571--2575, 2012.

\bibitem{hwang76b}
F.~K.~Hwang, ``Group Testing with a Dilution Effect," \emph{Biometrika}, vol.~63, no.~3, pp.~671--680, 1976.

\bibitem{laarhoven12wifs}
T.~Laarhoven, J.-J.~Oosterwijk, and J.~Doumen, ``Dynamic Traitor Tracing for Arbitrary Alphabets: Divide and Conquer," \emph{IEEE Workshop on Information Forensics and Security (WIFS)}, pp.~240--245, 2012.

\bibitem{laarhoven13tit}
T.~Laarhoven, J.~Doumen, P.~Roelse, B.~\v{S}kori\'{c}, and B.~de~Weger, ``Dynamic Tardos Traitor Tracing Schemes," \emph{IEEE Transactions on Information Theory}, vol.~59, no.~7, pp.~4230--4242, 2013.

\bibitem{laarhoven13allerton}
T.~Laarhoven, ``Efficient Probabilistic Group Testing Based on Traitor Tracing," \emph{51st Annual Allerton Conference on Communication, Control and Computing (Allerton)}, pp.~1458--1465, 2013.

\bibitem{laarhoven13wifs}
T.~Laarhoven, ``Dynamic Traitor Tracing Schemes, Revisited," \emph{IEEE Workshop on Information Forensics and Security (WIFS)}, pp.~191--196, 2013.

\bibitem{laarhoven14ihmmsec}
T.~Laarhoven, ``Capacities and Capacity-Achieving Decoders for Various Fingerprinting Games," \emph{ACM Workshop on Information Hiding and Multimedia Security (IH\&MMSec)}, 2014. This is a preliminary version of the present paper.

\bibitem{laarhoven14decoders}
T.~Laarhoven, ``Asymptotics of Fingerprinting and Group Testing: Capacity-Achieving Log-Likelihood Decoders," \emph{submitted to IEEE Transactions on Information Theory}, 2014.

\bibitem{lebedev10}
V.~S.~Lebedev, ``Separating Codes and a New Combinatorial Search Model," \emph{Problems of Information Transmission}, vol.~46, no.~1, pp.~1--6, 2010.

\bibitem{meerwald11b}
P.~Meerwald and T.~Furon, ``Group Testing Meets Traitor Tracing," \emph{IEEE International Conference on Acoustics, Speech and Signal Processing (ICASSP)}, pp.~4204--4207, 2011.

\bibitem{meerwald12}
P.~Meerwald and T.~Furon, ``Toward Practical Joint Decoding of Binary Tardos Fingerprinting Codes," \emph{IEEE Transactions on Information Forensics and Security}, vol.~7, no.~4, pp.~1168--1180, 2012.

\bibitem{moulin08}
P.~Moulin, ``Universal Fingerprinting: Capacity and Random-Coding Exponents," \emph{arXiv:0801.3837v3 [cs.IT]}, 2011.


\bibitem{oosterwijk13b}
J.-J.~Oosterwijk, B.~\v{S}kori\'{c}, and J.~Doumen, ``A Capacity-Achieving Simple Decoder for Bias-Based Traitor Tracing Schemes," \emph{Cryptology ePrint Archive}, 2013.

\bibitem{oosterwijk14}
J.-J.~Oosterwijk, J.~Doumen, and T.~Laarhoven, ``Tuple Decoders for Traitor Tracing Schemes," \emph{SPIE Proceedings}, vol.~9028, 2014.

\bibitem{perez09}
L.~P\'{e}rez-Freire and T.~Furon, ``Blind Decoder for Binary Probabilistic Traitor Tracing Codes," \emph{IEEE Workshop on Information Forensics and Security (WIFS)}, pp.~46--50, 2009.

\bibitem{sebo85}
A.~Seb\H{o}, ``On Two Random Search Problems," \emph{Journal of Statistical Planning and Inference}, vol.~11, pp.~23--31, 1985.

\bibitem{sejdinovic10}
D.~Sejdinovic and O.~Johnson, ``Note on Noisy Group Testing: Asymptotic Bounds and Belief Propagation Reconstruction," \emph{48th Allerton Conference on Communication, Control, and Computing (Allerton)}, pp.~998--1003, 2010.

\bibitem{skoric08} 
B.~\v{S}kori\'{c}, S.~Katzenbeisser, and M.~U.~Celik, ``Symmetric Tardos Fingerprinting Codes for Arbitrary Alphabet Sizes," \emph{Designs, Codes and Cryptography}, vol.~46, no.~2, pp.~137--166, 2008.

\bibitem{stinson00}
D.~R.~Stinson, T.~van~Trung, and R.~Wei, ``Secure Frameproof Codes, Key Distribution Patterns, Group Testing Algorithms and Related Structures," \emph{Journal of Statistical Planning and Inference}, vol.~86, no.~2, pp.~595--617, 2000.

\bibitem{tardos03}
G.~Tardos, ``Optimal Probabilistic Fingerprint Codes," \emph{35th ACM Symposium on Theory of Computing (STOC)}, pp.~116--125, 2003.

\end{thebibliography}
\end{document}